\pdfoutput=1
\RequirePackage{ifpdf}
\ifpdf 
\documentclass[pdftex]{sigma}
\else
\documentclass{sigma}
\fi

\numberwithin{equation}{section}

\usepackage{tikz}
\usetikzlibrary{arrows}
\usetikzlibrary{positioning,decorations.text}

\begin{document}

\allowdisplaybreaks

\newcommand{\arXivNumber}{1801.00960}

\renewcommand{\thefootnote}{}

\renewcommand{\PaperNumber}{036}

\FirstPageHeading

\ShortArticleName{Surface Defects in E-String Compactifications and the van Diejen Model}

\ArticleName{Surface Defects in E-String Compactifications\\ and the van Diejen Model\footnote{This paper is a~contribution to the Special Issue on Elliptic Hypergeometric Functions and Their Applications. The full collection is available at \href{https://www.emis.de/journals/SIGMA/EHF2017.html}{https://www.emis.de/journals/SIGMA/EHF2017.html}}}

\Author{Belal NAZZAL and Shlomo S.~RAZAMAT}
\AuthorNameForHeading{B.~Nazzal and S.S.~Razamat}
\Address{Department of Physics, Technion, Haifa 32000, Israel}
\Email{\href{mailto:sban@campus.technion.ac.il}{sban@campus.technion.ac.il}, \href{mailto:razamat@physics.technion.ac.il}{razamat@physics.technion.ac.il}}

\ArticleDates{Received January 11, 2018, in final form April 03, 2018; Published online April 17, 2018}

\Abstract{We study the supersymmetric index of four dimensional theories obtained by compactifications of the six dimensional E string theory on a Riemann surface. In particular we derive the difference operator introducing certain class of surface defects to the index computation. The difference operator turns out to be, up to a constant shift, an analytic difference operator discussed by van Diejen.}


\Classification{81T60}

\renewcommand{\thefootnote}{\arabic{footnote}}
\setcounter{footnote}{0}

\section{Introduction}\label{section1}

Embedding four dimensional supersymmetric quantum field theories into a six dimensional model compactified on a Riemann surface can lead to various insights into the physics in four dimensions. Some examples of such insights include geometrization of the understandings of the space of theories in four dimensions. In particular, this leads to a geometrical understandings of conformal manifolds, deformations, symmetries, and dualities of the theories in four dimensions. Another example is that of relating supersymmetric observables in models realized by compactification to lower dimensional physics, such as two dimensional conformal field theories, topological models, and quantum mechanical integrable models.

An interesting aspect of the latter insights is that the computations of some of the supersymmetric observables in different four dimensional models, labeled by choices made upon compacti\-fication such as the six dimensional model and the two dimensional geometry, are related to lower dimensional computation in models which are labeled only by the type of computation in four dimensions and the choice of the six dimensional model. Here the choice of geometry enters as the choice of the computation one performs in the two dimensional models. One example is the AGT \cite{Alday:2009aq} correspondence between sphere partition functions of ${\mathcal N}=2$ theories obtained by compactifications on a surface of $(2,0)$ theories, and computations of correlation functions in Liouville--Toda models. The choice of the particular model in two dimensions is determined by a choice of $(2,0)$ theory and the choice of correlation function is determined by compactification geometry.

Yet another example is that of the supersymmetric index and the relation of it to integrable quantum mechanical models \cite{Gadde:2009kb,Gadde:2011uv,Gaiotto:2012xa}.\footnote{The quantum mechanical integrable models make appearance in various contexts when studying supersymmetric theories in four dimensions \cite{Derkachov:2012iv,Gorsky:1993pe,Nekrasov:2015wsu,Nekrasov:2009rc,Yaamjr}.} The $(2,0)$ models in six dimensions are labeled by a~choice of ADE algebra. Taking one of these models in six dimensions, the index of theories in four dimensions is closely related to the ADE type Ruijsenaars--Schneider model. In particular, the Hamiltonians of such models, when acting on the parameters of the index associated to punctures on the Riemann surface, introduce defect operators into the index \cite{Gaiotto:2012xa}. This leads to the expectation that determining the eigen-functions of the integrable model one can compute index of any theory resulting in the compactifications by combining them in a way determined by the geometry \cite{Gadde:2011uv}. At the technical level the Hamiltonians appear as one computes certain residues of the index of models with no defects and claiming that these correspond to indices of other models with defects \cite{Gaiotto:2012xa}. For similar derivation of difference operators corresponding to Hamiltonians of integrable systems see~\cite{Derkachov:2012iv}.

The correspondence between integrable models and compactifications can be generalized to situations where the six dimensional theory is a more generic $(1,0)$ model; leading to models in four dimensions with ${\mathcal N}=1$ supersymmetry. The supersymmetric index is well defined for such four dimensional theories and as it does not depend on continuous parameters it should define a topological invariant of the Riemann surface on which the $(1,0)$ is compactified. Moreover, the models in four dimensions can admit supersymmetric surface defects. The defects can be engineered by giving a space-time dependent vacuum expectation values to certain chiral operator and following the flow to the new infra-red fixed point \cite{Gaiotto:2012xa}. The index of this fixed point is obtained by computing a residue of the index of the UV theory \cite{Gaiotto:2012xa}. The residue calculus for the theories obtained by compactifications implies that the index of a~theory with a~surface defect can be obtained from the index of the theory without a~defect by acting on it with a~difference operator. As mentioned above, the difference operator in the case of $(2,0)$ theories is the Ruijsenaars--Schneider analytic difference operator. If one considers type A M5 branes on type $A_{k-1}$ singularity we obtain a generalization of such operators \cite{Gaiotto:2015usa,Ito:2016fpl,Maruyoshi:2016caf,Yagi:2017hmj}.\footnote{See~\cite{Arthamonov:2017oxw} for appearance of similar looking operators in a different setting.} We can consider a more general setup with the models in six dimensions labeled by a pair of algebras $(G,\hat G)$ with the first one denoting the choice of the M5 branes and the second the choice of singularity. In particular $(ADE,A_0)$ are the Ruijsenaars--Schneider models and the $(A,A)$ were computed in~\cite{Gaiotto:2015usa,Ito:2016fpl,Maruyoshi:2016caf,Yagi:2017hmj}.

In this note we show that the integrable model $(A_0,D_4)$ corresponding to one M5 brane on~$D_4$ singularity, also know as the E-string, is the $BC_1$ van Diejen model~\cite{vd}. What allows us to have this computation done for the E-string theory is the fact that we know rather explicitly the indices of all the models in this class of theories~\cite{Kim:2017toz}.

The E-string model has $E_8$ symmetry in six dimensions and the theories in four dimensions have symmetry which is some sub-group of~$E_8$. At the level of the difference operator the $E_8$ structure is hidden with only ${\rm U}(1)\times {\rm SU}(8)$ appearing. A~reason for this is that the difference operators act on symmetries associated to punctures and the punctures break the $E_8$ symmetry to this group~\cite{Kim:2017toz}. However, the symmetry of theories with no punctures can enhance to larger groups and in particular to $E_8$. The $E_8$ structure has been observed also when studying the difference operator in its own right, see~\cite{rseh}.

\looseness=-1 For other types of compactifications, such as A type M5 branes probing ADE singularity, we know some of the theories but not enough at the moment to have a derivation of the operators. However, the general program of relating compactifications to integrable models suggests that there might be integrable models labeled by some choices of a pair of ADE groups or in more generality on a choice of $(1,0)$ theory. The $(1,0)$ theories recently underwent classification attempts \cite{Bhardwaj:2015xxa,DelZotto:2014hpa, GAiotto:2014lca,Heckman:2015bfa, Heckman:2013pva,Morrison:2012np}. A vast variety of such models exists and the ones giving rise to four dimensional models with weakly coupled gauge fields in four dimensions, which is indicated by a gauge description once theory is compactified on a circle to five dimensions, might be relevant for deriving new difference operators. It will be very enlightening to understand this structure.

The note is organized as following. We begin with the review of the index ingredients we need for the computation for the E-string theories. In section three we discuss the computation of the operator introducing surface defects. In section four we relate the operator to the van Diejen model. In section five we discuss a limit of the operator for which the eigen-functions are known to be given in terms of Koornwinder polynomials. We have few appendices with more technical details of the computations.

\section{E-string compactifications}\label{section2}

Let us review the essential details of the compactification program for the case of one M5 brane probing $D_4$ singularity, the E string theory, and introduce the basic building blocks of the construction~\cite{Kim:2017toz}. We compactify the E string theory on a Riemann surface with punctures in presence of flux for abelian sub-groups of the $E_8$ global symmetry supported on the surface. The states of the six dimensional theory have to have good transformation properties in presence of the flux and thus it needs to be properly quantized. We associate models in four dimensions then to punctured Riemann surfaces and to choice of flux. The models have a subgroup of $E_8$ as the symmetry. The subgroup is determined by the choice of the flux and punctures, that is it is the subgroup of $E_8$ commuting with the choice of flux in the case of closed Riemann surface. For properly quantized flux this has rank eight, for fractional values of flux the rank might be smaller. Every puncture is associated with additional factor of ${\rm SU}(2)$ symmetry. The punctures come in different types which we refer to as different colors. We will continue the discussion in the language of the index as it encodes all the needed physical information. For definitions see Appendix~\ref{appendixA}.

Models corresponding to different surfaces can be glued together by gauging a symmetry corresponding to punctures of the same color. The color of the punctures determines what are the details of the gluing. The punctures break the $E_8$ symmetry of the six dimensional model to ${\rm U}(1)\times {\rm SU}(8)$ sub group. The flux might break the symmetry further. In particular the color is determined by the ${\rm U}(1)\times {\rm SU}(8)$ subgroup of $E_8$ which the puncture keeps. The subgroup preserved by given puncture is parametrized by fugacity $t$ for ${\rm U}(1)$ and fugacities $a_i$ for ${\rm SU}(8)$ ($i=1,\dots, 8$ and $\prod\limits_{i=1}^8 a_i=1$). For different colors of punctures the fugacities of one are expressible in terms of monomial products of the other. When we glue two punctures together the index of the theory is
\begin{gather*}
T_{\rm combined}=T_{\mathfrak J}^{A}(u)\times_u T_{\mathfrak J}^{B}(u)\equiv (q;q)(p;p)\oint\frac{{\rm d}u}{4\pi i u} \frac{\prod\limits_{j=1}^8\Gamma_e\big((q p)^{\frac12 }\frac1{t^{\mathfrak J}} \big(a^{\mathfrak J}_j\big)^{-1} u^{\pm1}\big)}{\Gamma(u^{\pm2})} T_{\mathfrak J}^{A}(u)T_{\mathfrak J}^{B}(u) .
\end{gather*}
Here the indices $A$ and $B$ stand for {\it Theory A} and {\it Theory B}. The gamma functions appearing in the denominator correspond to ${\mathcal N}=1$ vector fields and the gamma functions in the numerator to a collection of eight chiral fields in fundamental representation of the gauged symmetry. This collection of chiral fields couples to certain chiral operators of the two glued copies which generalize the moment map operators of the class~${\mathcal S}$ case~\cite{Kim:2017toz}.

We will use the shorthand notation $\times_u$ to indicate the gluing. Here $T_{\mathfrak J}(u)$ is an index of a theory corresponding to some Riemann surface with puncture of color ${\mathfrak J}$ with associated symmetry ${\rm SU}(2)_u$. The parameters $t^{\mathfrak J}$ and $a^{\mathfrak J}$ label the ${\rm U}(1)\times {\rm SU}(8)$ symmetry preserved by the puncture.

Let us define the basic building blocks of our construction. We define the tube $T_{{\mathfrak J},\overline{\mathfrak J}}(z,u)$ to be
\begin{gather*}
T_{{\mathfrak J},\overline{\mathfrak J}}(u,z)=\Gamma_e\big(q p t^4\big)\left(\prod_{j=1}^8\Gamma_e\big((q p)^{\frac12}t a_j z^{\pm1}\big)\Gamma_e\big(( q p)^{\frac12} t a_j^{-1} u^{\pm1}\big)\right)\Gamma_e\left(\frac1{t^2}u^{\pm1}z^{\pm1}\right) .
\end{gather*}
This tube is the model obtained as compactification on sphere with two punctures and flux $-1/2$ for ${\rm U}(1)_t$ and zero flux for other symmetries. The model is an IR free Wess--Zumino theory. From this we can construct cap theories $C^{(M,L; i)}_{{\mathfrak J}}(z)$, corresponding to a sphere with single puncture, by computing residues~\cite{Gaiotto:2012xa}. We define these to be
\begin{gather*}
C^{(M,L;i)}_{{\mathfrak J}}(z)=\frac1{\Big(\prod\limits_{j\neq i}\Gamma_e(a_i/a_j)\Big)\Gamma_e\big( pq t^2\frac1{a_i^2}\big)(q;q)(p;p)}\operatorname{Res}_{u\to \frac1{(q p)^{\frac12} q^M p^L t} a_i} \frac1u T_{{\mathfrak J},\overline{\mathfrak J}}(u,z) .
\end{gather*}
 The cap theory for zero values of $M$ and $L$ is a model corresponding to sphere with one puncture and flux $-\frac34$ for ${\rm U}(1)_t$, $\frac78$ for ${\rm U}(1)_i$, and $-\frac18$ for ${\rm U}(1)_j$. See~\cite{Kim:2017toz} for details of the derivation of the flux.
The index can be thought as partition function on ${\mathbb S}^1\times {\mathbb S}^3$, and for non vanishing values of~$M$ and~$L$ the theory
also has surface defects wrapping the~${\mathbb S}^1$ and one of the two equators of~${\mathbb S}^3$. Finally we have a three punctured sphere $T_{{\mathfrak J}_B,{\mathfrak J}_C,{\mathfrak J}_D}(w,u,v)$
\begin{gather*}
T_{{\mathfrak J}_B,{\mathfrak J}_C,{\mathfrak J}_D}(w,u,v) = \Gamma_e\big((q p)^{\frac12} t\big(B^{-1}A\big)^{\pm1} w^{\pm1}\big)\Gamma_e\left(\frac{q p}{t^2}\right) (q;q)(p;p)\\
\quad{} \times \oint \frac{{\rm d} h}{4\pi i h} \frac{\Gamma_e\big(\frac{(p q)^{\frac12}}{t^2}\big(A B^{-1}\big)^{\pm1}h^{\pm1}\big)}{\Gamma_e\big(h^{\pm2}\big)}\Gamma_e\big(t h^{\pm1} w^{\pm1}\big) H\big(u,D,v,C,\sqrt{h B},\sqrt{h^{-1}B}; A\big),
\end{gather*} where we have defined
\begin{gather}
H(z_1,z_2,v_1,v_2, a , b; A ) = (q;q)^2(p;p)^2\oint\frac{{\rm d}w_1}{4\pi i w_1}\oint \frac{{\rm d}w_2}{4\pi i w_2} \frac{\Gamma_e\big(\frac{(p q)^{\frac12}}{t^2}w_1^{\pm1}w_2^{\pm1}\big)}{\Gamma_e\big(w_2^{\pm2}\big)\Gamma_e\big(w_1^{\pm2}\big)} \nonumber\\
 \quad{} \times \Gamma_e\big( (q p)^{\frac14}t A^{\frac12}b^{-1} w_1^{\pm1}z_1^{\pm1}\big) \Gamma_e\big((qp)^{\frac14} A^{\frac12} bw_1^{\pm1}z_2^{\pm1}\big)\Gamma_e\big((q p)^{\frac14} t A^{-\frac12} b w_2^{\pm1} z_1^{\pm1}\big)\nonumber\\
\quad{} \times \Gamma_e \big((q p)^{\frac14} A^{-\frac12} b^{-1} z_2^{\pm1}w_2^{\pm1}\big) \Gamma_e\big((q p)^{\frac14} t A^{-\frac12} a w_1^{\pm1} v_1^{\pm1}\big)\Gamma_e\big( ( q p)^{\frac14} A^{-\frac12} a^{-1} v_2^{\pm1} w_1^{\pm1}\big)\nonumber\\
\quad{} \times \Gamma_e\big( ( q p )^{\frac14} t A^{\frac12} a^{-1} w_2^{\pm1} v_1^{\pm1}\big) \Gamma_e\big( ( q p )^{\frac14} A^{\frac12} a w_2^{\pm1} v_2^{\pm1}\big) . \label{gff}
 \end{gather}

The above expressions are non trivial to derive. The theory corresponding to three punctured spheres is constructed by starting from a gauge theory, index of which is roughly speaking $H$, and arguing that at some point on the conformal manifold the ${\rm U}(1)$ symmetry corresponding to fugacity $\sqrt{a/b}$ enhances to ${\rm SU}(2)$. This is a non trivial fact which follows from dualities. This~${\rm SU}(2)$ is then taken to be dynamical with addition of some chiral fields. The resulting index is given above. The statement that this theory corresponds to three punctured sphere is made by performing a variety of physical consistency checks~\cite{Kim:2017toz}. Note that the construction also gives a theory having only rank five symmetry as opposed to rank eight.

For the three punctured sphere we have flux $3/4$ for ${\rm U}(1)_t$ and vanishing flux for the Cartan generators of ${\rm SU}(8)$. The three punctured sphere depends on four parameters $(A,B,C,D)$ which parametrize ${\rm SO}(8)$ inside ${\rm SU}(8)$. That is,
\begin{gather}\label{aj}
 (a_1,a_2,a_3,a_4) =A^{\pm1} B^{\pm1} ,\qquad (a_5,a_6,a_7,a_8) =C^{\pm1} D^{\pm1} .
\end{gather}
In principle there should be three punctured spheres depending on all eight parameters but the particular construction of~\cite{Kim:2017toz} gives us a three punctured sphere only depending on five with the map to eight parameters written above.

The three punctures are of different color
\begin{gather*}
w\colon \ {\mathfrak J}_B=\big(t; A^{\pm1}B^{\pm1},C^{\pm1}D^{\pm1}\big) , \\
u\colon \ {\mathfrak J}_C=\big(t;A^{\pm1}D^{\pm1},B^{\pm1}C^{\pm1}\big) , \\
v\colon \ {\mathfrak J}_D=\big(t; A^{\pm1}C^{\pm1}, B^{\pm1}D^{\pm1}\big) .
\end{gather*}
Without loss of any generality let us assume that we will compute residues with respect to $a_1=A B^{-1}$. Then as we have only a subgroup of ${\rm SU}(8)$ we need to specify the flux for this. We obtain that the flux for the cap is
\begin{gather*} \big({\rm U}(1)_A,{\rm U}(1)_B,{\rm U}(1)_C,{\rm U}(1)_D\big) = \big(\tfrac14,-\tfrac14,0,0\big) .
\end{gather*}

\begin{figure}[t]\centering \small
\begin{tikzpicture}
\node at (0,0){\includegraphics[scale=0.74]{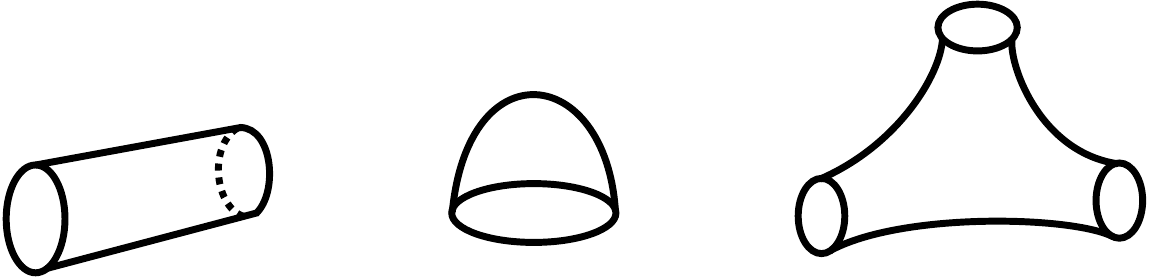}};

\node at (1.4,-0.8){${\mathfrak J}_B$};

\node at (4.6,-0.8){${\mathfrak J}_D$};

\node at (3.0,1.3){${\mathfrak J}_C$};

\node at (-0.3,0.6){$(M,L;i)$};

\node at (-0.3,-1.0){${\mathfrak J}$};

\node at (-4.5,-0.6){${\mathfrak J}$};

\node at (-2.1,-0.1){$\overline{\mathfrak J}$};

\end{tikzpicture}
\caption{The three basic ingredients. The tube on the left, the cap, and the three punctured sphere. The punctures can be of various colors determined by an octet of variables ${\mathfrak J}$ and the cap is labelled by the residue used for its definition.}\label{Fig1}
\end{figure}
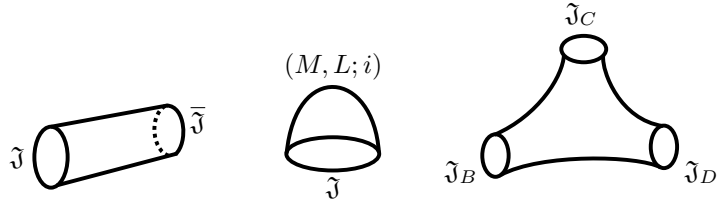

\section{Defect operators}\label{section3}

Using the building blocks of the previous section we can introduce surface defects into the index computation. Given a model of some flux and corresponding to some surface we introduce a~defect operator by gluing to the surface first two three punctured spheres and then closing two of the punctures with caps. In case one closes the two punctures with cap defined by residues $(0,0;i)$ and $(0,0;\overline i)$, where by $\overline i$ we mean $a_j$ such that $a_i=1/a_j$, one adds tube with zero flux, which gives us the original model without defect. We can indeed check, see Appendix~\ref{appendixB}, that the index satisfies such property
\begin{gather}
T_{{\mathfrak J}_C}(u)= T_{{\mathfrak J}_C}(z)\times_z \big(\big(T_{{\mathfrak J}_B, {\mathfrak J}_C,{\mathfrak J}_D}(h,z,g)\times_h C^{(0,0;i)}_{{\mathfrak J}_B}(h)\big)\nonumber\\
\hphantom{T_{{\mathfrak J}_C}(u)=}{}
\times_g\big(T_{{\mathfrak J}_B, {\mathfrak J}_C,{\mathfrak J}_D}(v,u,g)\times_v C^{(0,0;\overline i)}_{{\mathfrak J}_B}(v)\big)\big) .\label{tryuiop}
\end{gather}

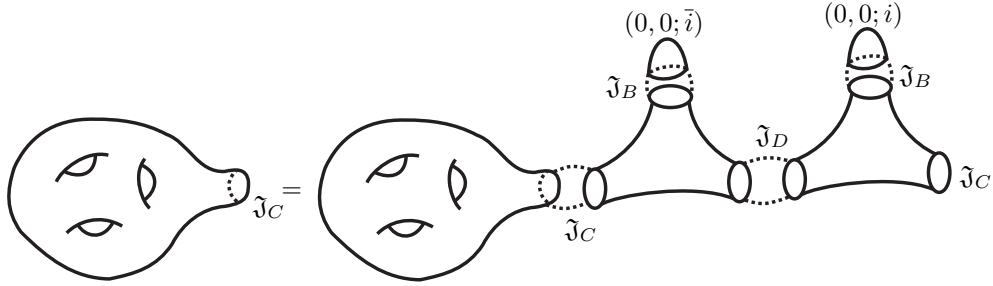
\begin{figure}[t]\centering\small
\begin{tikzpicture}
\node at (0,0){\includegraphics[scale=0.65]{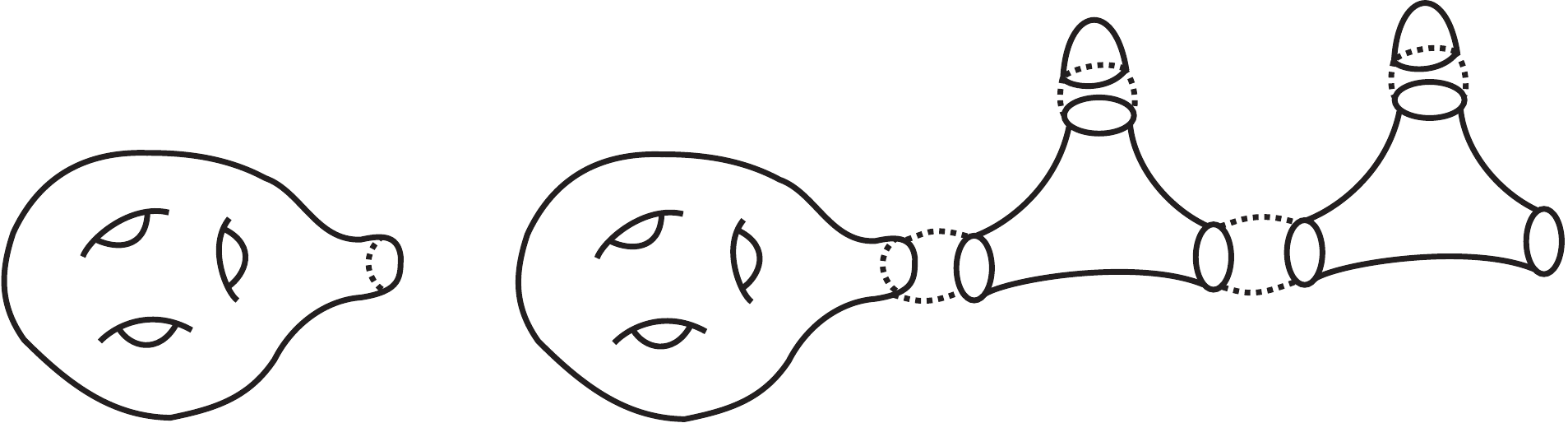}};

\node at (1.35,-1.0){${\mathfrak J}_C$};

\node at (6.6,-0.3){${\mathfrak J}_C$};

\node at (-2.8,-0.7){${\mathfrak J}_C$};

\node at (-2.5,-0.5){$=$};

\node at (3.9,0.25){${\mathfrak J}_D$};

\node at (1.9,0.9){${\mathfrak J}_B$};

\node at (5.8,1.05){${\mathfrak J}_B$};

\node at (5.1,1.85){$(0,0;i)$};

\node at (2.45,1.75){$(0,0;\overline{i})$};

\end{tikzpicture}
\caption{Gluing two three punctured spheres with two punctures closed as indicated gives the original model.}\label{Fig2}
\end{figure}

\begin{figure}[t!]\centering\small
\begin{tikzpicture}
\node at (0,0){\includegraphics[scale=0.65]{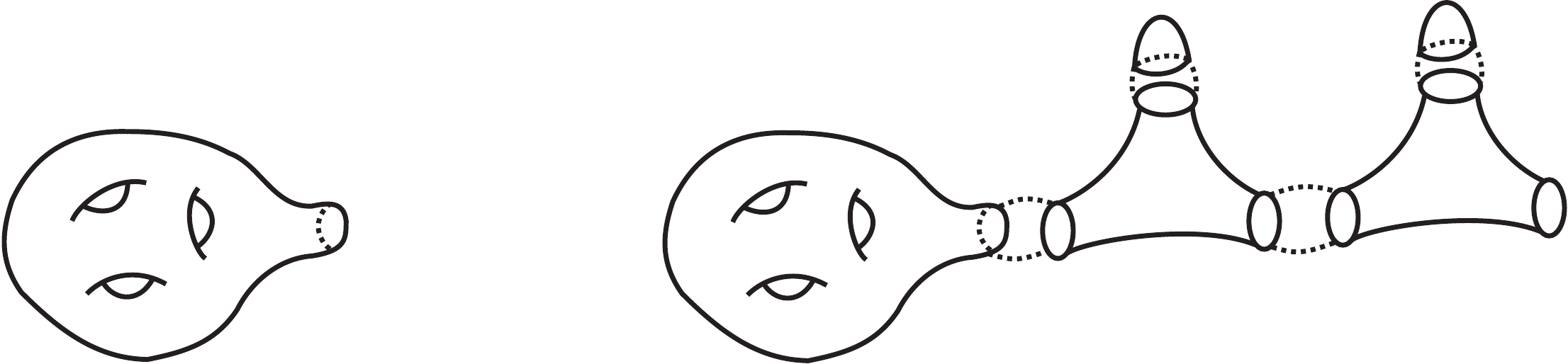}};

\node at (2.3,-1.0){${\mathfrak J}_D$};

\node at (7.6,-0.3){${\mathfrak J}_D$};

\node at (-3.8,-0.7){${\mathfrak J}_D$};

\node at (-2.4,-0.5){$\times D_{{\mathfrak J}_D}^{{\mathfrak J}_B,(1,0;i)}=$};

\node at (4.8,0.2){${\mathfrak J}_C$};

\node at (2.95,1.0){${\mathfrak J}_B$};

\node at (6.8,1.1){${\mathfrak J}_B$};

\node at (6.2,1.85){$(1,0;i)$};

\node at (3.5,1.75){$(0,0;\overline{i})$};

\end{tikzpicture}
\caption{Gluing two three punctured spheres with two punctures closed as indicated gives the original model with surface defect, and for the index we act on the index of model with no defect by a difference operator.}\label{Fig3}
\end{figure}
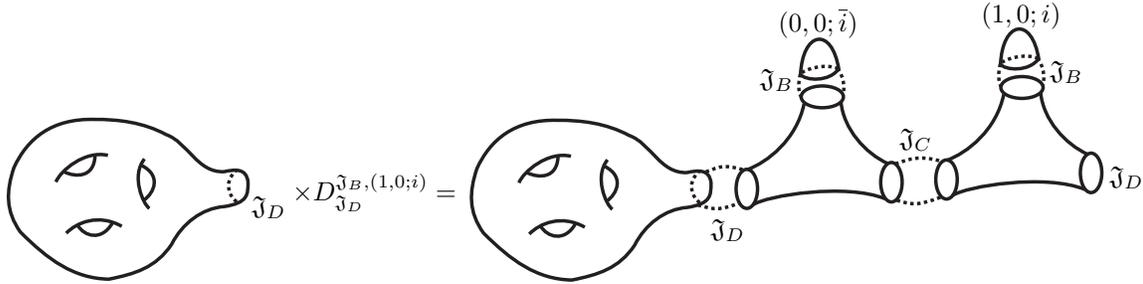

However, when we close one of the punctures with $(M,L;i)$ and other with $(0,0;\overline i)$ we introduce a surface defect. Performing the computation with $M=1$ and $L=0$, see Appendix~\ref{appendixC}, we can see that the index is given by acting on the one with no defect by a~difference operator
\begin{gather*}
{\mathfrak D}_{{\mathfrak J}_D}^{{\mathfrak J}_B,(1,0;i)}T_{{\mathfrak J}_D}(u)=
T_{{\mathfrak J}_D}(g)\times_g \big(\big(T_{{\mathfrak J}_B, {\mathfrak J}_C,{\mathfrak J}_D}(h,z,g)\times_h C^{(0,0;i)}_{{\mathfrak J}_B}(h)\big)\\
\hphantom{{\mathfrak D}_{{\mathfrak J}_D}^{{\mathfrak J}_B,(1,0;i)}T_{{\mathfrak J}_D}(u)=}{}
\times_z\big(T_{{\mathfrak J}_B, {\mathfrak J}_C,{\mathfrak J}_D}(v,z,u)\times_v C^{(1,0;\overline i)}_{{\mathfrak J}_B}(v)\big)\big) .\nonumber
\end{gather*}

The difference operator is given by
\begin{gather}
{\mathfrak D}_{{\mathfrak J}_D}^{{\mathfrak J}_B,(1,0;AB^{-1})} T_{{\mathfrak J}_D}(z) \sim
\frac{\theta_p\big((pq)^\frac{1}{2}t^{-1}A^{\pm1}C^{\pm1}z\big)\theta_p\big((pq)^\frac{1}{2}t^{-1}B^{\pm1}D^{\pm1}z\big)}{\theta_p\big(q z^2\big)\theta_p\big(z^2\big)} T_{{\mathfrak J}_D}(qz)\nonumber\\
\hphantom{{\mathfrak D}_{{\mathfrak J}_D}^{{\mathfrak J}_B,(1,0;AB^{-1})} T_{{\mathfrak J}_D}(z) \sim}{}
 +\frac{\theta_p\big((pq)^\frac{1}{2}t^{-1}A^{\pm1}C^{\pm1}z^{-1}\big)\theta_p((pq)^\frac{1}{2}t^{-1}B^{\pm1}D^{\pm1}z^{-1})}{\theta_p\big(q z^{-2}\big)\theta_p\big(z^{-2}\big)} T_{{\mathfrak J}_D}\big(qz^{-1}\big)\nonumber\\
\hphantom{{\mathfrak D}_{{\mathfrak J}_D}^{{\mathfrak J}_B,(1,0;AB^{-1})} T_{{\mathfrak J}_D}(z) \sim}{}
+ W^{{\mathfrak J}_B}_{{\mathfrak J}_D,\big(1,0;AB^{-1}\big)}(z) T_{{\mathfrak J}_D}(z) ,\label{diuer}
\end{gather}
where $\sim$ means equal up to an overall factor which is independent of~$z$. We have denoted
\begin{gather*}
W^{{\mathfrak J}_B}_{{\mathfrak J}_D, (1,0; AB^{-1})}(z)
\\=\frac{\theta_p\big(q^{-1}t^{-4}\big)\theta_p\big(q^{-1}t^{-4}A^2B^{-2}z^2\big)\theta_p\big((pq)^\frac{1}{2}t^{\pm1}A C^{\pm1}(qz)^{-1}\big)\theta_p\big((pq)^\frac{1}{2}t^{\pm1}B^{-1}D^{\pm1}(qz)^{-1}\big)}
{\theta_p\big(q^{-2}t^{-4}A^2B^{-2}\big)\theta_p\big(z^2\big)\theta_p\big(q^{-1}z^{-2}\big)\theta_p\big(t^{-4}z^2\big)}\\
\quad{} +\frac{\theta_p\big(q^{-1}t^{-4}\big)\theta_p\big(q^{-1}t^{-4}A^2B^{-2}z^{-2}\big)\theta_p\big((pq)^\frac{1}{2}t^{\pm1}A^{-1} C^{\pm1}z^{-1}\big)\theta_p\big((pq)^\frac{1}{2}t^{\pm1}BD^{\pm1}z^{-1}\big)}
{\theta_p\big(q^{-2}t^{-4}A^2B^{-2}\big)\theta_p\big(z^{-2}\big)\theta_p\big(q^{-1}z^{2}\big)\theta_p\big(t^{-4}z^{-2}\big)}\\
\quad{} +\frac{\theta_p\big(q^{-1}A^2B^{-2}\big)\theta_p\big((pq)^\frac{1}{2}t^{2}B D^{\pm1}\big(t^{-1}z\big)^{\pm1}\big)
\theta_p\big((pq)^\frac{1}{2}t^{2}A^{-1} C^{\pm1} \big(t^{-1} z\big)^{\pm1}\big)}
{\theta_p\big(q^{-2}t^{-4}A^2B^{-2}\big)\theta_p\big(z^2\big)\theta_p\big(t^4z^{-2}\big)} \\
\quad{}+\frac{\theta_p\big(q^{-1}A^2B^{-2}\big)\theta_p\big((pq)^\frac{1}{2}t^{2}B D^{\pm1}\big(t^{-1}z^{-1}\big)^{\pm1}\big)
\theta_p\big((pq)^\frac{1}{2}t^{2}A^{-1} C^{\pm1} \big(t^{-1} z^{-1}\big)^{\pm1}\big)}
{\theta_p\big(q^{-2}t^{-4}A^2B^{-2}\big)\theta_p\big(z^{-2}\big)\theta_p\big(t^4z^2\big)}\\
\quad{} +\frac{\theta_p\big(t^{-2}\big)\theta_p\big(q^{-1}t^{-2}A^2\big)\theta_p\big(q^{-1}A^2B^{-2}\big)\theta_p\big(q^{-1}t^{-2}B^{-2}\big) \theta_p\big(q^{-1}t^{-2}A B^{-1} C^{\pm1} D^{\pm1}\big)}{\theta_p\big(p^{-1}q^{-2}t^{-4}A^2 B^{-2}\big)\theta_p\big(q^{-2}t^{-2}A^2B^{-2}\big)}.
\end{gather*}
In Appendix~\ref{appendixC} we give details of the computation leading to this operator. One could consider more general residues by gluing the cap $C^{(L,M;\overline i)}_{{\mathfrak J}}(v)$ and general~$L$ and~$M$. We leave this as an exercise to the interested reader.

\section{Relation to van Diejen model}\label{section4}

The difference operator of the previous section is the van Diejen difference operator. Using the notations of \cite{Reji} and the definitions of Appendix~\ref{appendixA} the van Diejen operator is given as
\begin{gather*}
A_D(h;z)T(z)\equiv V(h;z)T(qz)+V\big(h;z^{-1}\big)T\big(q^{-1}z\big)+V_b(h;z) ,
\end{gather*}
where
\begin{gather*}
 V(h;z) \equiv \frac{\prod\limits_{n=1}^8\theta\big((pq)^{\frac12}h_n z\big)}{\theta(z^2)\theta\big(qz^2\big)},\qquad
 V_b(h;z) \equiv \frac{\sum\limits_{k=0}^3p_k(h)[\mathcal{E}_k(\xi;z)-\mathcal{E}_k(\xi;\omega_k)]}{2\theta(\xi)\theta\big(q^{-1}\xi\big)} ,
\end{gather*}
where $\omega_k$ are
\begin{gather*}
 \omega_0=1 , \qquad \omega_1=-1 , \qquad \omega_2=p^\frac12 ,\qquad \omega_3=-p^{-\frac12} .
\end{gather*}
The functions $p_k(h)$ are
\begin{alignat*}{3}
 & p_0(h)\equiv \prod_n \theta(p^\frac12 h_n) , \qquad && p_1(h)\equiv \prod_n \theta\big({-}p^\frac12 h_n\big) ,& \\
 & p_2(h)\equiv p\prod_n h_n^{-\frac12}\theta(h_n) ,\qquad && p_3(h)\equiv p\prod_n h_n^{\frac12}\theta\big({-}h_n^{-1}\big) ,&
\end{alignat*}
and $\mathcal{E}_k$ is
\begin{gather*}
 \mathcal{E}_k(\xi;z)\equiv\frac{\theta\big(q^{-\frac12} \xi \omega_k^{-1} z\big)\theta\big(q^{-\frac12} \xi \omega_k z^{-1}\big)}{\theta\big(q^{-\frac12}\omega_k^{-1} z\big)\theta\big(q^{-\frac12} \omega_k z^{-1}\big)} .
\end{gather*}

The van Diejen operator and the operator (\ref{diuer}) are the same up to a constant function (independent of~$z$). It's clear that $V(h;z)$ coincides with the corresponding term in \eqref{diuer} if we make the identifications
\begin{gather*}
 h_{1,2,3,4}=t^{-1}A^{\pm1}C^{\pm1} , \qquad h_{5,6,7,8}=t^{-1}B^{\pm1}D^{\pm1} .
\end{gather*}
Since $V_b(h;z)$ is elliptic in $z$ with periods~$1$ and $p$ and it is easy to check that $W^{{\mathfrak J}_B}_{{\mathfrak J}_D, (1,0; AB^{-1})}(z)$ is also elliptic with the same period, it is enough to show that the two functions have the same poles and residues to prove that they can differ only by a function independent of~$z$. In the fundamental parallelogram $V_b$ has poles at (we assume with no loss of generality that $|p|<|q|\ll |t|<1$ and the rest of the variables are on unit circle)
\begin{gather*}
 z=\pm q^{-\frac12}p , \qquad z= \pm q^{\frac12}, \qquad z=\pm p^\frac12 q^{\pm\frac12}.
\end{gather*}
In addition to such poles the operator \eqref{diuer} seems to have poles at $z=\pm t^{-2}p , \pm t^2 , 	\pm p^\frac12 t^{\pm2}$ and $z=\pm1,\pm p^\frac12$, but computation of the residue at these poles yields zero. The computation of the residue at the poles is straightforward, the result is ($h$ is either $1$ or $-1$)
\begin{gather*}
\operatorname{Res}_{z\to h q^\frac12} W^{{\mathfrak J}_B}_{{\mathfrak J}_D, (1,0; AB^{-1})}(z)=
-h(p;p)^{-2}\frac{\theta_p\big(h p^\frac{1}{2}t^{\pm1}A C^{\pm1}\big)\theta_p\big(h p^\frac{1}{2}t^{\pm1}B^{-1}D^{\pm1}\big)}
{2q^{-\frac12}\theta_p\big(q^{-1}\big)},\\
\operatorname{Res}_{z\to h q^{-\frac12}} W^{{\mathfrak J}_B}_{{\mathfrak J}_D, (1,0; AB^{-1})}(z)=
h(p;p)^{-2}\frac{\theta_p\big(hp^\frac{1}{2}t^{\pm1}A C^{\pm1}\big)\theta_p\big(h p^\frac{1}{2}t^{\pm1}B^{-1}D^{\pm1}\big)}
{2q^{\frac12}\theta_p\big(q^{-1}\big)},\\
\operatorname{Res}_{z\to h p^\frac12 q^{\frac12}} W^{{\mathfrak J}_B}_{{\mathfrak J}_D, (1,0; AB^{-1})}(z)=
-h (p;p)^{-2}\frac{A^{-2}B^{2}\theta_p\big(h t^{\pm1}A C^{\pm1}\big)\theta_p\big(h t^{\pm1}B^{-1}D^{\pm1}\big)}
{2p^{-\frac32}q^{-\frac12}\theta_p\big(q^{-1}\big)},\\
\operatorname{Res}_{z\to h p^\frac12 q^{-\frac12}} W^{{\mathfrak J}_B}_{{\mathfrak J}_D, (1,0; AB^{-1})}(z)=
h (p;p)^{-2}\frac{A^{-2}B^{2}\theta_p\big(h t^{\pm1}A C^{\pm1}\big)\theta_p\big(h t^{\pm1}B^{-1}D^{\pm1}\big)}
{2p^{-\frac32}q^{\frac12}\theta_p\big(q^{-1}\big)}.
\end{gather*}
Using the ellipticity of $V_b$ and some basic properties of the theta function this is exactly what we get from computing the residues of $V_b$. Thus, we can conclude that the operators are the same up to a constant function. The van Diejen operator does not depend on the type of residue we took, that is what type of defect was introduced, but only on the color of puncture through the choice of the eight parameters. The choice of the defect enters through the additive constant by which the operator derived from the index differs from the van Diejen operator. In particular, operators introducing different defects commute with each other as they differ by a constant. We can compute operators which correspond to residues with $q$ exchanged by $p$, these will correspond to defects wrapping the other equator of ${\mathbb S}^3$. All the operators should commute and indeed they do as this is true for the van Diejen operators. Our derivation of the operator had only five parameters but the relation discussed here suggests generalization to eight parameters, again up to the additive constant function.

\section{Koornwinder limit}\label{section5}

We consider the following limit of the parameters. Define
\begin{gather}\label{limpfuj}
p^{\frac12} \widetilde A =A C , \qquad p^{\frac12} \widetilde C= B D \qquad \widetilde B=A/C ,\qquad \widetilde D=B/D .
\end{gather}
We take $p$ to zero keeping the new variables fixed. In the limit the difference operator is
\begin{gather*}
{\mathfrak D}_{{\mathfrak J}_D}^{{\mathfrak J}_B,\big(1,0;(\widetilde A \widetilde B/ \widetilde C \widetilde D)^{\frac12}\big)} T_{{\mathfrak J}_D}(u)\\
{} \sim \frac{\big(1-q^{-\frac12}t \widetilde A^{-1}u^{-1}\big)\big(1-q^{-\frac12}t \widetilde C^{-1}u^{-1}\big)\big(1-q^{\frac12} t^{-1} \widetilde A^{-1} u\big)\big(1-q^{\frac12} t^{-1} \widetilde C^{-1} u\big)}{\big(1-u^2\big)\big(1-q u^2\big)} T_{{\mathfrak J}_D}(q u) \\
 \quad{}+\frac{\big(1-q^{-\frac12}t \widetilde A^{-1}u\big)\big(1-q^{-\frac12}t \widetilde C^{-1}u\big)\big(1-q^{\frac12} t^{-1} \widetilde A^{-1} u^{-1}\big)\big(1-q^{\frac12} t^{-1} \widetilde C^{-1} u^{-1}\big)}{\big(1-u^{-2}\big)\big(1-q u^{-2}\big)} T_{{\mathfrak J}_D}(q u^{-1}) \\
\quad{} +W^{{\mathfrak J}_B}_{{\mathfrak J}_D,\big(1,0;(\widetilde A \widetilde B/\widetilde C\widetilde D )^{\frac12}\big)} T_{{\mathfrak J}_D}(u).\end{gather*}
 We note that conjugating the operator to be
\begin{gather*}
{\mathfrak O}_{{\mathfrak J}_D}^{{\mathfrak J}_B,\big(1,0;(\widetilde A \widetilde B / \widetilde C \widetilde D)^{\frac12}\big)}=
\Gamma_e\big(q^{\frac12}t {\widetilde A}^{-1}u^{\pm1}\big)^{-1} \Gamma_e\big(q^{\frac12}t {\widetilde C}^{-1}u^{\pm1}\big)^{-1}\\
 \hphantom{{\mathfrak O}_{{\mathfrak J}_D}^{{\mathfrak J}_B,\big(1,0;(\widetilde A \widetilde B / \widetilde C \widetilde D)^{\frac12}\big)}= }{}
\times {\mathfrak D}_{{\mathfrak J}_D}^{{\mathfrak J}_B,\big(1,0;(\widetilde A \widetilde B / \widetilde D
\widetilde C)^{\frac12}\big)} \Gamma_e\big(q^{\frac12}t {\widetilde A}^{-1}u^{\pm1}\big) \Gamma_e\big(q^{\frac12}t {\widetilde C}^{-1}u^{\pm1}\big) . \end{gather*}
We can write
\begin{gather*}
{\mathfrak O}_{{\mathfrak J}_D}^{{\mathfrak J}_B,\big(1,0;(\widetilde A \widetilde B/ \widetilde D \widetilde C)^{\frac12}\big)} {\mathfrak F}(u)=
\frac{\prod\limits_{l=1}^4(1- a_l u)}{\big(1-u^2\big)\big(1-q u^2\big)}({\mathfrak F}(q u)-{\mathfrak F}(u))\\
\qquad{}
+\frac{\prod\limits_{j=1}^4\big(1- a_j u^{-1}\big)}{\big(1-u^{-2}\big)\big(1-q u^{-2}\big)}\big({\mathfrak F}\big(q u^{-1}\big)-{\mathfrak F}(u)\big)
 +E_{{\mathfrak J}_D}^{{\mathfrak J}_B,\big(1,0;(\widetilde A \widetilde B / \widetilde D \widetilde C)^{\frac12}\big)} {\mathfrak F}(u) ,\\
a_1=q^{\frac12}t \widetilde A^{-1} ,\qquad
a_2=q^{\frac12}t \widetilde C^{-1} ,\qquad
a_3=q^{\frac12}t^{-1} \widetilde A^{-1} ,\qquad
a_4=q^{\frac12}t^{-1} \widetilde C^{-1}.
\end{gather*} The first two terms define the rank one Koornwinder operator \cite{koour} and we have an additional constant term. The eigenfunctions are the Askey--Wilson polynomials. In general these polynomials have four independent parameters $a_l$ but in our case they are restricted to $a_1 a_4 =a_2 a_3$. The constant term is
\begin{gather*}
E_{{\mathfrak J}_D}^{{\mathfrak J}_B,(1,0;\big(\widetilde A \widetilde B / \widetilde D
\widetilde C)^{\frac12}\big)}=\left[-\frac{q^3t^2}{ (\widetilde A \widetilde C)^2}-q^2t^2\left(1-\frac{\widetilde B}{\widetilde C\widetilde D\widetilde A}-\frac1{\widetilde C^2}-\frac{1+t^{-2}}{\widetilde A\widetilde C}\right)\right. \\
\left.\hphantom{E_{{\mathfrak J}_D}^{{\mathfrak J}_B,(1,0;\big(\widetilde A \widetilde B / \widetilde D \widetilde C)^{\frac12}\big)}=}{}
 - q\left(-\frac{\widetilde B}{\widetilde C^3\widetilde A\widetilde D}+\frac{\widetilde B}{\widetilde C^2\widetilde D}+\frac{\widetilde B}{\widetilde A\widetilde D\widetilde C}+\frac1{\widetilde C^2}+\frac{\widetilde B t^2}{\widetilde C^2 \widetilde D}\right)+\frac{\widetilde B\widetilde A}{\widetilde D \widetilde C}\right]\frac1{\frac{\widetilde A\widetilde B}{\widetilde C \widetilde D}-q^2t^2}.
\end{gather*}
Let us take the limit of the three punctured sphere. We give details of the computation in Appendix~\ref{appendixD} with the final result given here (taking $\lim\limits_{p\to 0}$ of the right-hand side as in~\eqref{limpfuj})
\begin{gather}\label{krlty}
 T_{{\mathfrak J}_B,{\mathfrak J}_C,{\mathfrak J}_D}(w,u,v)= \frac1{\big(q p\frac1{ABC D}t^{-2};q\big)} \left(q p t^2 \frac1{A BC D};q\right)^2\left(q p \frac1{B A C D};q\right) \\
{} \times \frac1{\big(\sqrt{ q p} \frac{t}{AC} v^{\pm1},\sqrt{q p}\frac{t}{D B}v^{\pm1};q\big)}
\frac1{\big(\sqrt{q p} \frac{t}{AB}w^{\pm1},\sqrt{ q p} \frac{t}{D C} w^{\pm1};q\big)}\frac1{
\big(\sqrt{q p } \frac{t}{B C} u^{\pm1},\sqrt {q p } \frac{ t}{ A D } u^{\pm1};q\big)} . \nonumber
\end{gather} The index factorizes. In particular on general grounds \cite{Gadde:2011uv} we expect the index of the three punctured sphere to be
\begin{gather*}
\frac1{\big(\sqrt{ q p} \frac{t}{AC} v^{\pm1},\sqrt{q p}\frac{t}{D B}v^{\pm1};q\big)}
\frac1{\big(\sqrt{q p} \frac{t}{AB}w^{\pm1},\sqrt{ q p} \frac{t}{D C} w^{\pm1};q\big)}\\
\qquad{}\times \frac1{\big(\sqrt{q p } \frac{t}{B C} u^{\pm1},\sqrt {q p }\frac{ t}{ A D } u^{\pm1};q\big)}
 \sum_{\lambda} C_\lambda \psi_\lambda(u)\psi_\lambda(w)\psi_\lambda(v) ,
\end{gather*} where the sum is over all the eigenvalues of the Koornwinder operator and $\psi_\lambda(z)$ are Askey--Wilson polynomials. What we have shown above is that in the limit the three punctured sphere we have defined has all~$C_\lambda$ vanishing but the one corresponding to the constant polynomial. The Koornwinder polynomials have higher rank generalizations which should be relevant for higher rank E string theories. In those cases we do not know the three punctured spheres and the relation to Koornwinder polynomials can provide a useful tool to study the indices of these models. The limit we considered does not have a special physical meaning a priori, however the fact that the expressions become simple and the fact that one might generalize the discussion to the higher rank case, make the limit of potential interest.

\appendix

\section{Index definitions}\label{appendixA}

We compute the supersymmetric index \cite{Kinney:2005ej} using the standard definitions of \cite{Dolan:2008qi}. The index of chiral field charged under flavor ${\rm U}(1)$ symmetry with charge $S$ and having R-charge ${\mathfrak R}$ is
\begin{gather*}
 \Gamma_e\big((q p)^{\frac{ \mathfrak R}2} u^S\big) .
\end{gather*}
The parameter $u$ is fugacity for the flavor symmetry. We define here
\begin{gather*} \Gamma_e(u) = \prod_{i,j=0}^\infty \frac{1-\frac1u q^{i+1}p^{j+1}}{1- u p^i q^j} .
\end{gather*} We will use the following definitions
\begin{gather*}
 (s;q)= \prod_{i=1}^\infty \big(1-s q^{i-1}\big) ,\qquad \theta_r(u)=\prod_{j=1}^\infty \big(1- u r^{j-1}\big)\big(1-r^j/u\big) .
\end{gather*} Finally we use the condensed conventions
\begin{gather*}
f\big(y^{\pm1}\big) =f(1/y) f(y) ,	\qquad (s_1,\dots,s_k;q) =(s_1;q)\cdots(s_k;q) .
\end{gather*}
Contour integrals in the paper are around the unit circle unless we state otherwise.

\section{Computation of the sphere with two punctures}\label{appendixB}

We give here the derivation of equation \eqref{tryuiop}. The computation involves calculating several contour integrals over products of elliptic gamma functions and taking residues. In what follows we will present the computation in a condensed manner by first computing the caps and then gluing them to spheres with three punctures. As we will see this way of presenting the computation will be somewhat singular. A proper way to define the computation is first computing the integrals resulting in gluing together spheres with three punctures and tubes and then computing the residues, that is turning tubes to caps. The reader should think about the singular parts of the computation as done in this manner. We will use different relations between integrals of elliptic gamma functions which physically are manifestations of Seiberg dualities. The relevant identities are derived in~\cite{rainus, spir} and one can consult the review~\cite{resgu}.

We compute
\begin{gather*}
C^{(0,0;AB^{-1})}_{{\mathfrak J}_B}(w)\times_w T_{{\mathfrak J}_B,{\mathfrak J}_C,{\mathfrak J}_D}(w,u,v)
\\
=(q;q)(p;p)\oint\frac{{\rm d}w}{4\pi i w} \frac{\prod\limits_{j=1}^8\Gamma_e\big((q p)^{\frac12 }\frac1t a_j^{-1} w^{\pm1}\big)}{\Gamma\big(w^{\pm2}\big)} C^{(0,0;AB^{-1})}_{{\mathfrak J}_B}(w)T_{{\mathfrak J}_B,{\mathfrak J}_C,{\mathfrak J}_D}(w,u,v)
\\
=(q;q)^4(p;p)^4\!\oint\!\frac{{\rm d}w}{4\pi i w} \frac{\prod\limits_{j=1}^8\Gamma_e\big((q p)^{\frac12 }\frac1t a_j^{-1} w^{\pm1}\big)}{\Gamma\big(w^{\pm2}\big)}
\Gamma_e\big(pqt^4\big)\prod_{j\neq i}\Gamma_e\left(\frac{pqt^2}{a_ia_j}\right)
\prod_{j=1}^8\Gamma_e\big((pq)^{\frac{1}{2}}ta_jw^{\pm1}\big)\\
\quad{}\times\Gamma_e\left(\frac{(pq)^{\frac{1}{2}}w^{\pm1}}{ta_i}\right)\Gamma_e\left(\frac{a_iw^{\pm1}}{(pq)^{\frac{1}{2}}t^3}\right)
 \Gamma_e\big((q p)^{\frac12} t\big(B^{-1}A\big)^{\pm1} w^{\pm1}\big)\Gamma_e\left(\frac{q p}{t^2}\right) \\
\quad{}\times\oint \frac{{\rm d}y}{4\pi i y} \frac{\Gamma_e\big(\frac{(p q)^{\frac12}}{t^2}\big(A B^{-1}\big)^{\pm1}y^{\pm1}\big)}{\Gamma_e(y^{\pm2})}
\Gamma_e\big(t y^{\pm1} w^{\pm1}\big)\oint\frac{{\rm d}w_1}{4\pi i w_1}\oint \frac{{\rm d}w_2}{4\pi i w_2} \frac{\Gamma_e\big(\frac{(p q)^{\frac12}}{t^2}w_1^{\pm1}w_2^{\pm1}\big)}{\Gamma_e\big(w_2^{\pm2}\big)\Gamma_e\big(w_1^{\pm2}\big)} \\
\quad{}\times\Gamma_e\big( (q p)^{\frac14}t A^{\frac12}B^{-\frac12} y^{\frac12} w_1^{\pm1}u^{\pm1}\big)
\Gamma_e\big((qp)^{\frac14} A^{\frac12} B^{\frac12} y^{-\frac12} w_1^{\pm1}D^{\pm1}\big)\\
\quad{}\times \Gamma_e\big((q p)^{\frac14} t A^{-\frac12} B^{\frac12} y^{-\frac12} w_2^{\pm1} u^{\pm1}\big)
\Gamma_e \big((q p)^{\frac14} A^{-\frac12} B^{-\frac12} y^{\frac12} D^{\pm1}w_2^{\pm1}\big)\\
\quad{}\times
\Gamma_e\big((q p)^{\frac14} t A^{-\frac12} B^\frac12 y^\frac12 w_1^{\pm1} v^{\pm1}\big)\Gamma_e\big( ( q p)^{\frac14} A^{-\frac12} B^{-\frac12} y^{-\frac12} C^{\pm1} w_1^{\pm1}\big) \\
\quad{} \times\Gamma_e\big( ( q p )^{\frac14} t A^{\frac12} B^{-\frac12} y^{-\frac12} w_2^{\pm1} v^{\pm1}\big)\Gamma_e\big( ( q p )^{\frac14} A^{\frac12} B^\frac12 y^\frac12 w_2^{\pm1} C^{\pm1}\big) .
\end{gather*}
Plugging in the values of $a_j$ from \eqref{aj} where $a_i=AB^{-1}$ and using the identity $\Gamma_e\big(\frac{pq}{z}\big)\Gamma_e(z)=1$ we get
\begin{gather*}
(q;q)^4(p;p)^4\oint\frac{{\rm d}w}{4\pi i w} \frac{1}{\Gamma\big(w^{\pm2}\big)}
\Gamma_e\big(pqt^4\big)\Gamma_e\big(pqt^2\big)\Gamma_e\big(pqt^2B^2\big)\Gamma_e\big(pqt^2A^{-2}\big)\\
\quad{} \times\Gamma_e\big(pqt^2A^{-1}BC^{\pm1}D^{\pm1}\big) \Gamma_e\left(\frac{AB^{-1}w^{\pm1}}{(pq)^{\frac{1}{2}}t^3}\right)
 \Gamma_e\big((q p)^{\frac12} tA^{-1}B w^{\pm1}\big)\Gamma_e\left(\frac{q p}{t^2}\right)\\
 \quad{}\times \oint \frac{{\rm d}y}{4\pi i y} \frac{\Gamma_e\big(\frac{(p q)^{\frac12}}{t^2}\big(A B^{-1}\big)^{\pm1}y^{\pm1}\big)}{\Gamma_e\big(y^{\pm2}\big)}\Gamma_e\big(t y^{\pm1} w^{\pm1}\big)\\
\quad{} \times\oint\frac{{\rm d}w_1}{4\pi i w_1}\oint \frac{{\rm d}w_2}{4\pi i w_2} \frac{\Gamma_e\big(\frac{(p q)^{\frac12}}{t^2}w_1^{\pm1}w_2^{\pm1}\big)}{\Gamma_e\big(w_2^{\pm2}\big)\Gamma_e\big(w_1^{\pm2}\big)} \Gamma_e\big( (q p)^{\frac14}t A^{\frac12}B^{-\frac12} y^{\frac12} w_1^{\pm1}u^{\pm1}\big)\\
\quad{} \times \Gamma_e\big((qp)^{\frac14} A^{\frac12} B^{\frac12} y^{-\frac12} w_1^{\pm1}D^{\pm1}\big)\Gamma_e\big((q p)^{\frac14} t A^{-\frac12} B^{\frac12} y^{-\frac12} w_2^{\pm1} u^{\pm1}\big) \\
\quad{} \times \Gamma_e \big((q p)^{\frac14} A^{-\frac12} B^{-\frac12} y^{\frac12} D^{\pm1}w_2^{\pm1}\big)
 \Gamma_e\big((q p)^{\frac14} t A^{-\frac12} B^\frac12 y^\frac12 w_1^{\pm1} v^{\pm1}\big)\\
\quad{} \times \Gamma_e\big( ( q p)^{\frac14} A^{-\frac12} B^{-\frac12} y^{-\frac12} C^{\pm1} w_1^{\pm1}\big) \Gamma_e\big( ( q p )^{\frac14} t A^{\frac12} B^{-\frac12} y^{-\frac12} w_2^{\pm1} v^{\pm1}\big) \\
\quad{} \times \Gamma_e\big( ( q p )^{\frac14} A^{\frac12} B^\frac12 y^\frac12 w_2^{\pm1} C^{\pm1}\big) .
\end{gather*}
We can perform the integral over $w$ using the inversion formula~\cite{invspor} which sets $y=\frac{AB^{-1}}{(pq)^\frac12t^2}$, and we get
\begin{gather*}
(q;q)^2(p;p)^2\Gamma_e\big(pqt^2B^2\big)\Gamma_e\big(pqt^2A^{-2}\big)\Gamma_e\big(pqt^2A^{-1}BC^{\pm1}D^{\pm1}\big)
\Gamma_e(pq)\Gamma_e\big(t^{-4}A^2B^{-2}\big)\\
\quad{}\times \Gamma_e\big(pqA^{-2}B^2\big)
\oint\frac{{\rm d}w_1}{4\pi i w_1}\oint \frac{{\rm d}w_2}{4\pi i w_2} \frac{\Gamma_e\big(\frac{(p q)^{\frac12}}{t^2}w_1^{\pm1}w_2^{\pm1}\big)}{\Gamma_e\big(w_2^{\pm2}\big)\Gamma_e\big(w_1^{\pm2}\big)} \Gamma_e\big( AB^{-1} u^{\pm1} w_1^{\pm1}\big)\\
\quad {} \times \Gamma_e\big((qp)^{\frac12}t BD^{\pm1} w_1^{\pm1}\big)\Gamma_e\big((q p)^{\frac12} t^2 A^{-1} B u^{\pm1} w_2^{\pm1}\big) \Gamma_e \big(t^{-1} B^{-1} D^{\pm1}w_2^{\pm1}\big) \Gamma_e\big( v^{\pm1} w_1^{\pm1}\big)\\
\quad{} \times\Gamma_e\big((q p)^{\frac12}t A^{-1} C^{\pm1} w_1^{\pm1}\big) \Gamma_e\big( (q p)^{\frac12} t^2 v^{\pm1} w_2^{\pm1}\big) \Gamma_e\big( t^{-1} A C^{\pm1} w_2^{\pm1}\big) .
\end{gather*}
$\Gamma_e(pq)$ is zero but the integral over $w_1$ is pinched at $w_1=v^{\pm1}$ due to the term $\Gamma_e(v^{\pm1}w_1^{\pm1})$ such that the multiplication is finite and we get
\begin{gather*}
(q;q)(p;p) \Gamma_e\big(pqt^2B^2\big)\Gamma_e\big(pqt^2A^{-2}\big)\Gamma_e\big(pqt^2A^{-1}BC^{\pm1}D^{\pm1}\big)
\Gamma_e\big(t^{-4}A^2B^{-2}\big)\Gamma_e\big(pqA^{-2}B^2\big)\\
\quad{} \times\Gamma_e\big( AB^{-1} u^{\pm1} v^{\pm1}\big)\Gamma_e\big((qp)^{\frac12}t BD^{\pm1} v^{\pm1}\big)
\Gamma_e\big((q p)^{\frac12}t A^{-1} C^{\pm1} v^{\pm1}\big)\\
\quad{} \times\oint \frac{{\rm d}w_2}{4\pi i w_2} \frac{1}{\Gamma_e\big(w_2^{\pm2}\big)}
\Gamma_e\big((q p)^{\frac12} t^2 A^{-1} B u^{\pm1} w_2^{\pm1}\big) \Gamma_e \big(t^{-1} B^{-1} D^{\pm1}w_2^{\pm1}\big)
\Gamma_e\big( t^{-1} A C^{\pm1} w_2^{\pm1}\big) .
\end{gather*}
Integral over $w_2$ can be evaluated using the elliptic beta integral formula~\cite{spir}, and the final result is
\begin{gather*}
\Gamma_e\big(pqA^{-2}B^2\big) \Gamma_e\big( AB^{-1} u^{\pm1} v^{\pm1}\big)\Gamma_e\big((qp)^{\frac12}t BD^{\pm1} v^{\pm1}\big)
\Gamma_e\big((q p)^{\frac12}t A^{-1} C^{\pm1} v^{\pm1}\big)\\
\quad{}\times \Gamma_e\big((pq)^\frac12 t B C^{\pm1}u^{\pm1}\big) \Gamma_e\big((pq)^\frac12 t A^{-1}D^{\pm1}u^{\pm1}\big) .
\end{gather*}
We glue this to another three punctured sphere closed with $a_i=A^{-1}B$ and the claim is that gluing this to a given model we get the same model. Indeed we have
\begin{gather*}
 T_{{\mathfrak J}_C}(u)\times_u \big(\big(T_{{\mathfrak J}_B, {\mathfrak J}_C,{\mathfrak J}_D}(w,u,v)\times_w C^{(0,0;AB^{-1})}_{{\mathfrak J}_B}(w)\big)\\
 \quad{} \times_v\big(T_{{\mathfrak J}_B, {\mathfrak J}_C,{\mathfrak J}_D}(h,z,v)\times_h C^{(0,0;A^{-1}B)}_{{\mathfrak J}_B}(h)\big)\big)\\
=(q;q)^2(p;p)^2\Gamma_e\big(pq\big(A^{-2}B^2\big)^{\pm1}\big)\Gamma_e\big((pq)^\frac12 t B^{-1} C^{\pm1}z^{\pm1}\big)
\Gamma_e\big((pq)^\frac12 t AD^{\pm1}z^{\pm1}\big)\\
\quad{} \times\oint\frac{{\rm d}u}{4\pi i u} \frac{\Gamma_e\big((q p)^{\frac12 }t^{-1} A^{\pm1} D^{\pm1}u^{\pm1}\big)
\Gamma_e\big((q p)^{\frac12 }t^{-1} B^{\pm1} C^{\pm1}u^{\pm1}\big)}{\Gamma\big(u^{\pm2}\big)}\\
\quad{} \times\Gamma_e\big((pq)^\frac12 t B C^{\pm1}u^{\pm1}\big)
\Gamma_e\big((pq)^\frac12 t A^{-1}D^{\pm1}u^{\pm1}\big)T_{{\mathfrak J}_C}(u)\\
\quad{} \times\oint\frac{{\rm d}v}{4\pi i v} \frac{\Gamma_e\big((q p)^{\frac12 }t^{-1} A^{\pm1} C^{\pm1}v^{\pm1}\big)
\Gamma_e\big((q p)^{\frac12 }t^{-1} B^{\pm1} D^{\pm1}v^{\pm1}\big)}{\Gamma\big(v^{\pm2}\big)}\\
\quad{} \times\Gamma_e\big( AB^{-1} u^{\pm1} v^{\pm1}\big)\Gamma_e\big((qp)^{\frac12}t BD^{\pm1} v^{\pm1}\big)
\Gamma_e\big((q p)^{\frac12}t A^{-1} C^{\pm1} v^{\pm1}\big)\\
\quad{} \times\Gamma_e\big( A^{-1}B z^{\pm1} v^{\pm1}\big)\Gamma_e\big((qp)^{\frac12}t B^{-1}D^{\pm1} v^{\pm1}\big)\Gamma_e\big((q p)^{\frac12}t A C^{\pm1} v^{\pm1}\big) .
\end{gather*}
Terms cancel in the second integral and we get
\begin{gather*}
=(q;q)^2(p;p)^2\Gamma_e\big(pq\big(A^{-2}B^2\big)^{\pm1}\big)\Gamma_e\big((pq)^\frac12 t B^{-1} C^{\pm1}z^{\pm1}\big)
\Gamma_e\big((pq)^\frac12 t AD^{\pm1}z^{\pm1}\big)\\
\quad{}\times\oint\frac{{\rm d}u}{4\pi i u} \frac{\Gamma_e\big((q p)^{\frac12 }t^{-1} A^{\pm1} D^{\pm1}u^{\pm1}\big)
\Gamma_e\big((q p)^{\frac12 }t^{-1} B^{\pm1} C^{\pm1}u^{\pm1}\big)}{\Gamma\big(u^{\pm2}\big)}\\
\quad{}\times\Gamma_e\big((pq)^\frac12 t B C^{\pm1}u^{\pm1}\big)\Gamma_e\big((pq)^\frac12 t A^{-1}D^{\pm1}u^{\pm1}\big)T_{{\mathfrak J}_C}(u)\\
\quad{}\times\oint\frac{{\rm d}v}{4\pi i v} \frac{1}{\Gamma\big(v^{\pm2}\big)}
\Gamma_e\big( AB^{-1} u^{\pm1} v^{\pm1}\big) \Gamma_e\big( A^{-1}B z^{\pm1} v^{\pm1}\big) .
\end{gather*}
The integrals can be evaluated using the inversion formula which sets $u=z$:
\begin{gather*}
=\Gamma_e\big((pq)^\frac12 t B^{-1} C^{\pm1}z^{\pm1}\big)
\Gamma_e\big((pq)^\frac12 t AD^{\pm1}z^{\pm1}\big)\Gamma_e\big((q p)^{\frac12 }t^{-1} A^{\pm1} D^{\pm1}z^{\pm1}\big)\\
\quad{} \times\Gamma_e\big((q p)^{\frac12 }t^{-1} B^{\pm1} C^{\pm1}z^{\pm1}\big)\Gamma_e\big((pq)^\frac12 t B C^{\pm1}z^{\pm1}\big)
\Gamma_e\big((pq)^\frac12 t A^{-1}D^{\pm1}z^{\pm1}\big)T_{{\mathfrak J}_C}(z) .
\end{gather*}
We see that all terms cancel so we get
\begin{gather*}
 T_{{\mathfrak J}_C}(u)\times_u \big(\big(T_{{\mathfrak J}_B, {\mathfrak J}_C,{\mathfrak J}_D}(w,u,v)\times_w C^{(0,0;AB^{-1})}_{{\mathfrak J}_B}(w)\big)\\
\qquad{} \times_v\big(T_{{\mathfrak J}_B, {\mathfrak J}_C,{\mathfrak J}_D}(h,z,v)\times_h C^{(0,0;A^{-1}B)}_{{\mathfrak J}_B}(h)\big)\big)= T_{{\mathfrak J}_C}(z).
\end{gather*}

\section{Computation of the sphere with two punctures and a defect}\label{appendixC}

Here we compute the difference operator. The computation is a small twist on the one of the previous section, however it is less straightforward and thus we discuss it in detail. We compute
\begin{gather*}
T_{{\mathfrak J}_B, {\mathfrak J}_C,{\mathfrak J}_D}(w,u,v)\times_w C^{(1,0;AB^{-1})}_{{\mathfrak J}_B}(w)\\
=(q;q)(p;p)\oint\frac{{\rm d}w}{4\pi i w} \frac{\prod\limits_{j=1}^8\Gamma_e\big((q p)^{\frac12 }\frac1t a_j^{-1} w^{\pm1}\big)}{\Gamma\big(w^{\pm2}\big)} C^{(1,0;AB^{-1})}_{{\mathfrak J}_B}(w)T_{{\mathfrak J}_B,{\mathfrak J}_C,{\mathfrak J}_D}(w,u,v) \\
\sim\oint\frac{{\rm d}w}{4\pi i w} \frac{\prod\limits_{j=1}^8\Gamma_e\big((q p)^{\frac12 }\frac1t a_j^{-1} w^{\pm1}\big)}{\Gamma\big(w^{\pm2}\big)}
\prod_{j=1}^8\Gamma_e\big((pq)^{\frac{1}{2}}ta_jw^{\pm1}\big)
\Gamma_e\left(\frac{(pq)^\frac12 q w^{\pm1}}{tAB^{-1}}\right)\\
\quad{} \times\Gamma_e\left(\frac{AB^{-1}w^{\pm1}}{(pq)^{\frac{1}{2}} q t^3}\right)
 \Gamma_e\big((q p)^{\frac12} t\big(B^{-1}A\big)^{\pm1} w^{\pm1}\big)\Gamma_e\left(\frac{q p}{t^2}\right)
\oint \frac{{\rm d}y}{4\pi i y} \frac{\Gamma_e\big(\frac{(p q)^{\frac12}}{t^2}\big(A B^{-1}\big)^{\pm1}y^{\pm1}\big)}{\Gamma_e\big(y^{\pm2}\big)}\\
\quad{}\times\Gamma_e\big(t y^{\pm1} w^{\pm1}\big)\oint\frac{{\rm d}w_1}{4\pi i w_1}\oint \frac{{\rm d}w_2}{4\pi i w_2} \frac{\Gamma_e\big(\frac{(p q)^{\frac12}}{t^2}w_1^{\pm1}w_2^{\pm1}\big)}{\Gamma_e\big(w_2^{\pm2}\big)\Gamma_e\big(w_1^{\pm2}\big)} \Gamma_e\big( (q p)^{\frac14}t A^{\frac12}B^{-\frac12}y^{\frac12} w_1^{\pm1}u^{\pm1}\big)\\
\quad{}\times\Gamma_e\big((qp)^{\frac14} A^{\frac12} B^{\frac12}y^{-\frac12} w_1^{\pm1}D^{\pm1}\big)\Gamma_e\big((q p)^{\frac14} t A^{-\frac12} B^{\frac12} y^{-\frac12} w_2^{\pm1} u^{\pm1}\big)\\
\quad{} \times \Gamma_e \big((q p)^{\frac14} A^{-\frac12} B^{-\frac12} y^{\frac12} D^{\pm1}w_2^{\pm1}\big)\Gamma_e\big((q p)^{\frac14} t A^{-\frac12} B^\frac12 y^\frac12 w_1^{\pm1} v^{\pm1}\big)\\
\quad{} \times \Gamma_e\big( ( q p)^{\frac14} A^{-\frac12} B^{-\frac12} y^{-\frac12} C^{\pm1} w_1^{\pm1}\big) \Gamma_e\big( ( q p )^{\frac14} t A^{\frac12} B^{-\frac12} y^{-\frac12} w_2^{\pm1} v^{\pm1}\big)\\
\quad{}\times\Gamma_e\big( ( q p )^{\frac14} A^{\frac12} B^\frac12 y^\frac12 w_2^{\pm1} C^{\pm1}\big),
\end{gather*}
where $\sim$ means equality up to overall factors independent of $w$, $u$, $v$. Using the identity $\Gamma_e\big(\frac{pq}{z}\big)\Gamma_e(z)=1$ and the elliptic beta integral formula we evaluate the integral over~$w$ and up to overall factors we get
\begin{gather}
\Gamma_e\big(pq^2\big)\oint \frac{{\rm d}y}{4\pi i y} \frac{\Gamma_e\big((pq)^\frac12 q A^{-1} B y^{\pm1}\big)\Gamma_e\big((pq)^{-\frac12}q^{-1}t^{-2}AB^{-1}y^{\pm1}\big)}{\Gamma_e\big(y^{\pm2}\big)}\oint\frac{{\rm d}w_1}{4\pi i w_1}\oint \frac{{\rm d}w_2}{4\pi i w_2}\nonumber\\
\quad{} \times \frac{\Gamma_e\big(\frac{(p q)^{\frac12}}{t^2}w_1^{\pm1}w_2^{\pm1}\big)}{\Gamma_e\big(w_2^{\pm2}\big)\Gamma_e\big(w_1^{\pm2}\big)}\Gamma_e\big( (q p)^{\frac14}t A^{\frac12}B^{-\frac12} y^{\frac12} w_1^{\pm1}u^{\pm1}\big)\Gamma_e\big((qp)^{\frac14} A^{\frac12} B^{\frac12} y^{-\frac12} w_1^{\pm1}D^{\pm1}\big)
\nonumber\\
 \quad{} \times \Gamma_e\big((q p)^{\frac14} t A^{-\frac12} B^{\frac12} y^{-\frac12} w_2^{\pm1} u^{\pm1}\big) \Gamma_e \big((q p)^{\frac14} A^{-\frac12} B^{-\frac12} y^{\frac12} D^{\pm1}w_2^{\pm1}\big)\nonumber\\
 \quad{}\times \Gamma_e\big((q p)^{\frac14} t A^{-\frac12} B^\frac12 y^\frac12 w_1^{\pm1} v^{\pm1}\big)
 \Gamma_e\big( ( q p)^{\frac14} A^{-\frac12} B^{-\frac12} y^{-\frac12} C^{\pm1} w_1^{\pm1}\big)\nonumber\\
\quad{}\times \Gamma_e\big( ( q p )^{\frac14} t A^{\frac12} B^{-\frac12} y^{-\frac12} w_2^{\pm1} v^{\pm1}\big) \Gamma_e\big( ( q p )^{\frac14} A^{\frac12} B^\frac12 y^\frac12 w_2^{\pm1} C^{\pm1}\big) .\label{c2}
\end{gather}
We got a zero multiplying the integral, but we will see next that some of the integrals are pinched giving finite result. We start by evaluating the integral over $y$ using the residue theorem and we get that the integral over $w_1$ is pinched due to $\Gamma_e\big((q p)^{\frac14} t A^{-\frac12} B^\frac12 y^\frac12 w_1^{\pm1} v^{\pm1}\big)$ term for the poles $y_1=(pq)^{-\frac12}t^{-2}AB^{-1}$ and $y_2=(pq)^{-\frac12}q^{-1}t^{-2}AB^{-1}$. For $y_1$ the integral over $w_1$ is pinched at $w_1=v^{\pm1}$ and we proceed exactly as in appendix B to get the same result up to some overall factors
\begin{gather}
\frac12 \frac{\Gamma_e\big(qt^{-2}\big)\Gamma_e\big(pq^2t^2A^{-2}B^2\big)\Gamma_e\big(t^{-2}B^{-2}\big)\Gamma_e\big(t^{-2}A^2\big) \Gamma_e\big((pq)^{-1}q^{-1}t^{-4}A^2 B^{-2}\big)} {\Gamma_e\big((pq)^{-1}t^{-4}A^2 B^{-2}\big)}\nonumber\\
\quad{}\times\Gamma_e\big(t^{-2}AB^{-1}C^{\pm1}D^{\pm1}\big)
\Gamma_e\big( AB^{-1} u^{\pm1} v^{\pm1}\big)\Gamma_e\big((qp)^{\frac12}t BD^{\pm1} v^{\pm1}\big)
\Gamma_e\big((q p)^{\frac12}t A^{-1} C^{\pm1} v^{\pm1}\big)\nonumber\\
\quad{}\times\Gamma_e\big((pq)^\frac12 t B C^{\pm1}u^{\pm1}\big)\Gamma_e\big((pq)^\frac12 t A^{-1}D^{\pm1}u^{\pm1}\big).\label{c3}
\end{gather}
Now we look at the pole $y_2$. the integral over $w_1$ is pinched at $w_1=q^{\pm\frac12}v^{\pm1}$. Substituting these values \eqref{c2} now becomes
\begin{gather*}
\frac12\frac{\Gamma_e\big(t^{-2}\big)\Gamma_e\big(pq^3t^2A^{-2}B^2\big)}
{\Gamma_e\big(pq^3 t^4 A^{-2}B^2\big)}\frac{\Gamma_e\big(v^2\big)}{\Gamma_e\big(qv^2\big)}
\Gamma_e\big( q^{-\frac12}AB^{-1}u^{\pm1}\big(q^\frac12 v\big)^{\pm1}\big)\Gamma_e\big((qp)^{\frac12}q^\frac12 t B D^{\pm1}\big(q^\frac12 v\big)^{\pm1}\big)\\
\quad{} \times\Gamma_e\big( ( q p)^{\frac12} q^\frac12 t A^{-1} C^{\pm1} \big(q^\frac12 v\big)^{\pm1}\big)
\oint \frac{{\rm d}w_2}{4\pi i w_2} \frac{1}{\Gamma_e\big(w_2^{\pm2}\big)} \Gamma_e\left(\frac{(p q)^{\frac12}}{t^2}q^\frac12 v w_2^{\pm1}\right)\\
\quad{}\times \Gamma_e\big((q p)^{\frac12}q^\frac12 t^2 A^{-1} B w_2^{\pm1} u^{\pm1}\big) \Gamma_e \big(q^{-\frac12} t^{-1} B^{-1} D^{\pm1}w_2^{\pm1}\big)\\
\quad{}\times \Gamma_e\big( ( q p )^{\frac12}q^\frac12 t^2 w_2^{\pm1} v^{-1}\big) \Gamma_e\big( q^{-\frac12}t^{-1} A w_2^{\pm1} C^{\pm1}\big)
+\big\{v \leftrightarrow v^{-1} \big\} .
\end{gather*}
The integral here can be interpreted as index of the ${\rm SU}(2)$ gauge theory with four flavors. Using the Intriligator--Pouliot~\cite{Intriligator:1995ne} duality transformation (that is $V(\underline{s})={\prod\limits_{1\leq j < k\leq 8} \Gamma_e(s_j s_k)V(\sqrt{pq}/\underline{s})}$ in the notations of~\cite{resgu}) the expression becomes
\begin{gather*}
\frac12\frac{\Gamma_e\big(t^{-2}\big)\Gamma_e\big(pq^3t^2A^{-2}B^2\big)}
{\Gamma_e\big(pq^3 t^4 A^{-2}B^2\big)}\frac{\Gamma_e\big(v^2\big)}{\Gamma_e\big(qv^2\big)}
\Gamma_e\big( q^{-\frac12}AB^{-1}u^{\pm1}\big(q^\frac12 v\big)^{\pm1}\big)\Gamma_e\big((qp)^{\frac12}q^\frac12 t B D^{\pm1}\big(q^\frac12 v\big)^{\pm1}\big)\\
\quad{} \times\Gamma_e\big( ( q p)^{\frac12} q^\frac12 t A^{-1} C^{\pm1} \big(q^\frac12 v\big)^{\pm1}\big) \Gamma_e\big(pq^2A^{-1}Bvu^{\pm1}\big)\Gamma_e\big((pq)^\frac12 t^{-3}B^{-1}vD^{\pm1}\big)\\
\quad{} \times\Gamma_e\big((pq)^\frac12 t^{-3}AvC^{\pm1}\big)\Gamma_e\big(pq^2t^4A^{-2}B^2\big)\Gamma_e\big((pq)^\frac12 t A^{-1}u^{\pm1}D^{\pm1}\big)\Gamma_e\big((pq)^\frac12 tBu^{\pm1}C^{\pm1}\big)\\
\quad{} \times\Gamma_e\big(pq^2t^4A^{-1}B v^{-1}u^{\pm1}\big)\Gamma_e\big(q^{-1}t^{-2}B^{-2}\big)\Gamma_e\big(q^{-1}t^{-2}AB^{-1}C^{\pm1}D^{\pm1}\big)\\
\quad{}\times\Gamma_e\big((pq)^\frac12 t B^{-1}v^{-1}D^{\pm1}\big) \Gamma_e\big(q^{-1}t^{-2}A^2\big)\Gamma_e\big((pq)^\frac12 t Av^{-1}C^{\pm1}\big)\Gamma_e\big(pq^2\big)\\
\quad{}\times \oint \frac{{\rm d}w_2}{4\pi i w_2} \frac{1}{\Gamma_e\big(w_2^{\pm2}\big)} \Gamma_e\big(q^{-\frac12}t^2 v^{-1} w_2^{\pm1}\big) \Gamma_e\big(q^{-\frac12} t^{-2} A B^{-1} w_2^{\pm1} u^{\pm1}\big)\\
\quad{} \times\Gamma_e \big((pq)^{\frac12}q^\frac12 t B D^{\pm1}w_2^{\pm1}\big)
\Gamma_e\big( ( q p )^{\frac12}q^\frac12 t A^{-1} w_2^{\pm1} C^{-1}\big) \Gamma_e\big( q^{-\frac12}t^{-2} v w_2^{\pm1} \big)
+\big\{v \leftrightarrow v^{-1} \big\} .
\end{gather*}
We have zero multiplying the integral but the integral is pinched at $w_2=\big(q^{\pm\frac12}t^{-2}v\big)^{\pm1}$ due to the colliding of poles of the first and last elliptic gamma functions in the last integral. We get different contribution for each choice of sign of $q$. Substituting each one of the values of $w_2$ and using the identities $\Gamma_e(qz)=\theta_p(z)\Gamma_e(z)$ and $\Gamma_e\big(pq^2z^{-1}\big)\Gamma_e(z)=\theta_p\big(q^{-1}z\big)$ we get
\begin{gather*}
\frac12 \frac{\Gamma_e(t^{-2})\Gamma_e\big(pq^3t^2A^{-2}B^2\big)\Gamma_e\big(q^{-1}t^{-2}B^{-2}\big)\Gamma_e\big(q^{-1}t^{-2}B^2\big)}
{\theta_p(pq^2t^4A^{-2}B^2)}\\
\quad{}\times\Gamma_e\big(q^{-1}t^{-2}A B^{-1}C^{\pm1}D^{\pm1}\big)\Gamma_e\big((pq)^\frac12 tA^{-1} D^{\pm1}u^{\pm1}\big)\Gamma_e\big((pq)^\frac12 t B D^{\pm1}u^{\pm1}\big)\\
\quad{}\times\Gamma_e\big(q^{-\frac12}A B^{-1}u^{\pm1}\big(q^\frac12 v\big)^{\pm1}\big)\Gamma_e\big((pq)^\frac12 q^\frac12 t B D^{\pm1}\big(q^\frac12 v\big)^{\pm1}\big)\Gamma_e\big((pq)^\frac12 q^\frac12 t A^{-1} C^{\pm1} \big(q^\frac12 v\big)^{\pm1}\big) \\
\quad{}\times\frac{\theta_p\big(q^{-1}t^{-4}AB^{-1}vu^{\pm1}\big)\theta_p\big((pq)^\frac12 q^{-1} t B^{-1}v^{-1}D^{\pm1}\big)\theta_p\big((pq)^\frac12 q^{-1}t A v^{-1} C^{\pm1}\big)}{\theta_p\big(t^{-4}v^2\big)\theta_p\big(v^2\big)}\\
+ \frac12 \frac{\Gamma_e(t^{-2})\Gamma_e\big(pq^3t^2A^{-2}B^2\big)\Gamma_e\big(q^{-1}t^{-2}B^{-2}\big)\Gamma_e\big(q^{-1}t^{-2}B^2\big)}
{\theta_p\big(pq^2t^4A^{-2}B^2\big)}\\
\quad{}\times\Gamma_e\big(q^{-1}t^{-2}A B^{-1}C^{\pm1}D^{\pm1}\big)\Gamma_e\big((pq)^\frac12 tA^{-1} D^{\pm1}u^{\pm1}\big)\Gamma_e\big((pq)^\frac12 t B D^{\pm1}u^{\pm1}\big)\\
\quad{}\times\Gamma_e\big(A B^{-1}u^{\pm1} v^{\pm1}\big)\Gamma_e\big((pq)^\frac12 q^\frac12 t B D^{\pm1}\big(q^\frac12 v\big)^{\pm1}\big)\Gamma_e\big((pq)^\frac12 q^\frac12 t A^{-1} C^{\pm1} \big(q^\frac12 v\big)^{\pm1}\big) \\
\quad{}\times\frac{\theta_p\big((pq)^\frac12 q^{-1} t^{-3} B^{-1}vD^{\pm1}\big)\theta_p\big((pq)^\frac12 q^{-1}t^{-3} A v C^{\pm1}\big)}
{\theta_p\big(t^{4}v^{-2}\big)\theta_p\big(v^2\big)} + \big\{v \leftrightarrow v^{-1} \big\} .
\end{gather*}
Adding the contribution of \eqref{c3} and taking away overall factors the final result is
\begin{gather}
T_{{\mathfrak J}_B, {\mathfrak J}_C,{\mathfrak J}_D}(w,u,v)\times_w C^{(1,0;AB^{-1})}_{{\mathfrak J}_B}(w)\nonumber\\
\sim\frac{\theta_p\big(pq^2t^2A^{-2}B^2\big)
\theta_p\big((pq)^{-1}q^{-1}t^{-4}A^2B^{-2}\big)}
{\theta_p(t^{-2})\theta_p\big(q^{-1}t^{-2}B^{-2}\big)\theta_p\big(q^{-1}t^{-2}A^2\big)\theta_p\big(pq^2t^4A^{-2}B^2\big)
\theta_p\big(q^{-1}t^{-2}AB^{-1}C^{\pm1}D^{\pm1}\big)}\nonumber\\
\quad{}\times\frac{\theta_p\big(q^{-1}t^{-4}AB^{-1}vu^{\pm1}\big)
\theta_p\big((pq)^\frac{1}{2}q^{-1}tB^{-1}v^{-1}D^{\pm1}\big)
\theta_p\big((pq)^\frac{1}{2}q^{-1}tAv^{-1}C^{\pm1}\big)}
{\theta_p\big(t^{-4}v^2\big)\theta\big(v^2\big)}\nonumber\\
\quad{}\times\Gamma_e\big(q^{-\frac{1}{2}}A B^{-1}u^{\pm1}\big(q^\frac{1}{2}v\big)^{\pm1}\big)
\Gamma_e\big((pq)^\frac{1}{2}q^\frac{1}{2}t B D^{\pm1}\big(q^\frac{1}{2}v\big)^{\pm1}\big)
\Gamma_e\big((pq)^{\frac{1}{2}}q^\frac{1}{2}tA^{-1}C^{\pm1}(q^\frac{1}{2}v)^{\pm1}\big)
\nonumber\\
\quad{}\times\Gamma_e\big((pq)^\frac{1}{2}tA^{-1}u^{\pm1}D^{\pm1}\big)
\Gamma_e\big((pq)^\frac{1}{2}t B u^{\pm1}C^{\pm1}\big)\nonumber\\
\quad{}+\frac{\theta_p\big(pq^2t^2A^{-2}B^2\big)
\theta_p\big((pq)^{-1}q^{-1}t^{-4}A^2B^{-2}\big)}
{\theta_p\big(t^{-2}\big)\theta_p\big(q^{-1}t^{-2}B^{-2}\big)\theta_p\big(q^{-1}t^{-2}A^2\big)\theta_p\big(pq^2t^4A^{-2}B^2\big)
\theta_p\big(q^{-1}t^{-2}AB^{-1}C^{\pm1}D^{\pm1}\big)}\nonumber\\
\quad{}\times\Gamma_e\big((pq)^\frac{1}{2}t A^{-1}u^{\pm1}D^{\pm1}\big)
\Gamma_e\big((pq)^\frac{1}{2}t B u^{\pm1}C^{\pm1}\big)
\nonumber\\
\quad{}\times\Gamma_e\big(AB^{-1}u^{\pm1}v^{\pm1}\big)\Gamma_e\big((pq)^\frac{1}{2}q^\frac{1}{2}t B D^{\pm1}\big(q^\frac{1}{2}v\big)^{\pm1}\big)
\Gamma_e\big((pq)^{\frac{1}{2}}q^\frac{1}{2}tA^{-1}C^{\pm1}\big(q^\frac{1}{2}v\big)^{\pm1}\big)
\nonumber\\
\quad{}\times\frac{\theta_p\big((pq)^\frac{1}{2}q^{-1}t^{-3}B^{-1}v D^{\pm1}\big)\theta_p\big((pq)^\frac{1}{2}q^{-1}t^{-3}A v C^{\pm1}\big)}
{\theta_p\big(v^2\big)\theta_p\big(t^4v^{-2}\big)}+\big\{v\leftrightarrow v^{-1}\big\}\nonumber\\
 \quad{}+\Gamma_e\big(A B^{-1}u^{\pm1}v^{\pm1}\big)\Gamma_e\big((pq)^{\frac{1}{2}}t B D^{\pm1}v^{\pm1}\big)
\Gamma_e\big((pq)^{\frac{1}{2}}t A^{-1}C^{\pm1}v^{\pm1}\big)\nonumber\\
\quad{}\times
\Gamma_e\big((pq)^\frac{1}{2}tA^{-1}u^{\pm1}D^{\pm1}\big)\Gamma_e\big((pq)^\frac{1}{2}t B u^{\pm1}C^{\pm1}\big).\label{c7}
\end{gather}
Now we glue this as indicated in Fig.~\ref{Fig3}
\begin{gather*}
T_{{\mathfrak J}_D}(v)\times_v \big(\big(T_{{\mathfrak J}_B, {\mathfrak J}_C,{\mathfrak J}_D}(w,u,v)\times_w C^{(0,0;A^{-1}B)}_{{\mathfrak J}_B}(w)\big)\\
\quad{}\times_u\big(T_{{\mathfrak J}_B, {\mathfrak J}_C,{\mathfrak J}_D}(h,u,z)\times_h C^{(1,0;AB^{-1})}_{{\mathfrak J}_B}(h)\big)\big)\\
\sim \oint\frac{{\rm d}u}{4\pi i u} \frac{\Gamma_e\big((q p)^{\frac12 }t^{-1} A^{\pm1} D^{\pm1}u^{\pm1}\big)
\Gamma_e\big((q p)^{\frac12 }t^{-1} B^{\pm1} C^{\pm1}u^{\pm1}\big)}{\Gamma\big(u^{\pm2}\big)}\\
\quad{}\times\Gamma_e\big((pq)^\frac12 t B^{-1} C^{\pm1}u^{\pm1}\big)
\Gamma_e\big((pq)^\frac12 t A D^{\pm1}u^{\pm1}\big) \\
\quad{}\times\oint\frac{{\rm d}v}{4\pi i v} \frac{\Gamma_e\big((q p)^{\frac12 }t^{-1} A^{\pm1} C^{\pm1}v^{\pm1}\big)
\Gamma_e\big((q p)^{\frac12 }t^{-1} B^{\pm1} D^{\pm1}v^{\pm1}\big)}{\Gamma(v^{\pm2})}\\
\quad{}\times\Gamma_e\big( A^{-1}B u^{\pm1} v^{\pm1}\big)\Gamma_e\big((qp)^{\frac12}t B^{-1}D^{\pm1} v^{\pm1}\big)
\Gamma_e\big((q p)^{\frac12}t A C^{\pm1} v^{\pm1}\big)T_{{\mathfrak J}_D}(v)\\
\quad{}\times T_{{\mathfrak J}_B, {\mathfrak J}_C,{\mathfrak J}_D}(h,u,z)\times_h C^{(1,0;AB^{-1})}_{{\mathfrak J}_B}(h) .
\end{gather*}
Let's perform the computation for each term in \eqref{c7} separately. From the first term we get
\begin{gather*}
\frac{\theta_p\big(pq^2t^2A^{-2}B^2\big)
\theta_p\big((pq)^{-1}q^{-1}t^{-4}A^2B^{-2}\big)}
{\theta_p\big(t^{-2}\big)\theta_p\big(q^{-1}t^{-2}B^{-2}\big)\theta_p\big(q^{-1}t^{-2}A^2\big)\theta_p\big(pq^2t^4A^{-2}B^2\big)
\theta_p\big(q^{-1}t^{-2}AB^{-1}C^{\pm1}D^{\pm1}\big)}\\
\quad{}\times\oint\frac{{\rm d}u}{4\pi i u} \frac{\Gamma_e\big((q p)^{\frac12 }t^{-1} A^{\pm1} D^{\pm1}u^{\pm1}\big)
\Gamma_e\big((q p)^{\frac12 }t^{-1} B^{\pm1} C^{\pm1}u^{\pm1}\big)}{\Gamma\big(u^{\pm2}\big)}\\
\quad{}\times\Gamma_e\big((pq)^\frac12 t B^{-1} C^{\pm1}u^{\pm1}\big)
\Gamma_e\big((pq)^\frac12 t A D^{\pm1}u^{\pm1}\big)\nonumber\\
\quad{}\times\oint\frac{{\rm d}v}{4\pi i v} \frac{\Gamma_e\big((q p)^{\frac12 }t^{-1} A^{\pm1} C^{\pm1}v^{\pm1}\big)
\Gamma_e\big((q p)^{\frac12 }t^{-1} B^{\pm1} D^{\pm1}v^{\pm1}\big)}{\Gamma\big(v^{\pm2}\big)}\\
\quad{}\times\Gamma_e\big( A^{-1}B u^{\pm1} v^{\pm1}\big)\Gamma_e\big((qp)^{\frac12}t B^{-1}D^{\pm1} v^{\pm1}\big)
\Gamma_e\big((q p)^{\frac12}t A C^{\pm1} v^{\pm1}\big)T_{{\mathfrak J}_D}(v)\\
\quad{}\times\frac{\theta_p\big((pq)^\frac{1}{2}q^{-1}tB^{-1}z^{-1}D^{\pm1}\big)
\theta_p((pq)^\frac{1}{2}q^{-1}tAz^{-1}C^{\pm1})}
{\theta_p\big(t^{-4}z^2\big)\theta\big(z^2\big)}\\
\quad{}\times\Gamma_e\big(q^{-\frac{1}{2}}A B^{-1}u^{\pm1}\big(q^\frac{1}{2}z\big)^{\pm1}\big)
\Gamma_e\big((pq)^\frac{1}{2}q^\frac{1}{2}t B D^{\pm1}\big(q^\frac{1}{2}z\big)^{\pm1}\big)
\Gamma_e\big((pq)^{\frac{1}{2}}q^\frac{1}{2}tA^{-1}C^{\pm1}\big(q^\frac{1}{2}z\big)^{\pm1}\big)\\
\quad{}\times\Gamma_e\big(t^{-4}AB^{-1}zu^{\pm1}\big)\Gamma_e\big(pq^2 t^4 A^{-1}B z^{-1}u^{\pm1}\big)
\Gamma_e\big((pq)^\frac{1}{2}tA^{-1}u^{\pm1}D^{\pm1}\big)
\Gamma_e\big((pq)^\frac{1}{2}t B u^{\pm1}C^{\pm1}\big)\\
\quad{}+\big\{z\leftrightarrow z^{-1}\big\} .
\end{gather*}
Terms cancel and the integral over $u$ reduces to an integral over 6 gamma functions which can be evaluated as before using the elliptic beta integral formula. We get
\begin{gather*}
\frac{\theta_p\big(pq^2t^2A^{-2}B^2\big)\theta_p\big((pq)^{-1}q^{-1}t^{-4}A^2B^{-2}\big)}
{\theta_p\big(t^{-2}\big)\theta_p\big(q^{-1}t^{-2}B^{-2}\big)\theta_p\big(q^{-1}t^{-2}A^2\big)\theta_p\big(pq^2t^4A^{-2}B^2\big)
\theta_p\big(q^{-1}t^{-2}AB^{-1}C^{\pm1}D^{\pm1}\big)}\\
\quad{}\times\frac{\theta_p\big((pq)^\frac{1}{2}q^{-1}tB^{-1}z^{-1}D^{\pm1}\big)
\theta_p\big((pq)^\frac{1}{2}q^{-1}tAz^{-1}C^{\pm1}\big)
\Gamma_e\big((pq)^\frac{1}{2}q^\frac{1}{2}t B D^{\pm1}\big(q^\frac{1}{2}z\big)^{\pm1}\big)}{\theta_p\big(t^{-4}z^2\big)}\\
\quad{}\times\frac{\Gamma_e\big((pq)^{\frac{1}{2}}q^\frac{1}{2}tA^{-1}C^{\pm1}\big(q^\frac{1}{2}z\big)^{\pm1})}{\theta\big(z^2\big)}\\
\quad{}\times
\oint\frac{{\rm d}v}{4\pi i v} \frac{\Gamma_e\big((q p)^{\frac12 }t^{-1} A^{\pm1} C^{\pm1}v^{\pm1}\big)
\Gamma_e\big((q p)^{\frac12 }t^{-1} B^{\pm1} D^{\pm1}v^{\pm1}\big)}{\Gamma\big(v^{\pm2}\big)}\\
\quad{}\times\Gamma_e\big(zv^{\pm1}\big)\Gamma_e\big((qp)^{\frac12}t B^{-1}D^{\pm1} v^{\pm1}\big)
\Gamma_e\big((q p)^{\frac12}t A C^{\pm1} v^{\pm1}\big)
\Gamma_e\big(q^{-1}A^2B^{-2}\big)\Gamma_e\big(pq^2t^4\big)\\
\quad{}\times\Gamma_e\big(t^{-4}A^2B^{-2}z^2\big)\Gamma_e\big(pqt^4z^{-2}\big)
\Gamma_e\big(q^{-1}t^{-4}A^2B{-2}\big)\Gamma_e\big(q^{-1}z^{-1}v^{\pm1}\big)
\Gamma_e\big(pq^2\big)\\
\quad{}\times \Gamma_e\big(pq^2t^4A^{-2}B^2z^{-1}v^{\pm1}\big)\Gamma_e\big(t^{-4}zv^{\pm1}\big)
\Gamma_e\big(A^{-2}B^2\big)T_{{\mathfrak J}_D}(v) +\big\{z\leftrightarrow z^{-1}\big\} .
\end{gather*}
The integral over $v$ is pinched at $v=z^{\pm1},(qz)^{\pm1}$ due to collision of poles of $\Gamma_e\big(zv^{\pm1}\big)$ and $\Gamma_e\big(q^{-1}z^{-1}v^{\pm1}\big)$ and we get
\begin{gather*}
\frac{\theta_p(pq^2t^2A^{-2}B^2)\theta_p\big((pq)^{-1}q^{-1}t^{-4}A^2B^{-2}\big)}
{\theta_p\big(t^{-2}\big)\theta_p\big(q^{-1}t^{-2}B^{-2}\big)\theta_p\big(q^{-1}t^{-2}A^2\big)\theta_p\big(pq^2t^4A^{-2}B^2\big)
\theta_p\big(q^{-1}t^{-2}AB^{-1}C^{\pm1}D^{\pm1}\big)}\\
\times\frac{\theta_p\big((pq)^\frac{1}{2}q^{-1}tB^{-1}z^{-1}D^{\pm1}\big)
\theta_p\big((pq)^\frac{1}{2}q^{-1}tAz^{-1}C^{\pm1}\big)
\Gamma_e\big((pq)^\frac{1}{2}q^\frac{1}{2}t B D^{\pm1}\big(q^\frac{1}{2}z\big)^{\pm1}\big)}
{\theta_p\big(t^{-4}z^2\big)\theta\big(z^2\big)}\\
\times \frac{\Gamma_e\big((pq)^{\frac{1}{2}}q^\frac{1}{2}tA^{-1}C^{\pm1}\big(q^\frac{1}{2}z\big)^{\pm1}\big)\Gamma_e\big((q p)^{\frac12 }t^{-1} A^{\pm1} C^{\pm1}z^{\pm1}\big)
\Gamma_e\big((q p)^{\frac12 }t^{-1} B^{\pm1} D^{\pm1}z^{\pm1}\big)}{\Gamma\big(z^{\pm2}\big)}\\
\times\Gamma_e\big((qp)^{\frac12}t B^{-1}D^{\pm1} z^{\pm1}\big)
\Gamma_e\big((q p)^{\frac12}t A C^{\pm1} z^{\pm1}\big)
\Gamma_e\big(q^{-1}A^2B^{-2}\big)\Gamma_e\big(pq^2t^4\big)\Gamma_e\big(t^{-4}A^2B^{-2}z^2\big)\\
\times\Gamma_e\big(zz^{\pm1}\big)\Gamma_e\big(pqt^4z^{-2}\big)\Gamma_e\big(q^{-1}t^{-4}A^2B{-2}\big)\Gamma_e\big(q^{-1}z^{-1}z^{\pm1}\big)\\
\times\Gamma_e\big(pq^2\big)\Gamma_e\big(pq^2t^4A^{-2}B^2z^{-1}z^{\pm1}\big)\Gamma_e\big(t^{-4}zz^{\pm1}\big)
\Gamma_e\big(A^{-2}B^2\big)T_{{\mathfrak J}_D}(z)\\
+ \frac{\theta_p\big(pq^2t^2A^{-2}B^2\big)\theta_p\big((pq)^{-1}q^{-1}t^{-4}A^2B^{-2}\big)}
{\theta_p\big(t^{-2}\big)\theta_p\big(q^{-1}t^{-2}B^{-2}\big)\theta_p\big(q^{-1}t^{-2}A^2\big)\theta_p\big(pq^2t^4A^{-2}B^2\big)
\theta_p\big(q^{-1}t^{-2}AB^{-1}C^{\pm1}D^{\pm1}\big)}\\
\times\frac{\theta_p\big((pq)^\frac{1}{2}q^{-1}tB^{-1}z^{-1}D^{\pm1}\big)
\theta_p\big((pq)^\frac{1}{2}q^{-1}tAz^{-1}C^{\pm1}\big)
\Gamma_e\big((pq)^\frac{1}{2}q^\frac{1}{2}t B D^{\pm1}\big(q^\frac{1}{2}z\big)^{\pm1}\big)}
{\theta_p\big(t^{-4}z^2\big)\theta\big(z^2\big)}\\
\times\frac{\Gamma_e\big((pq)^{\frac{1}{2}}q^\frac{1}{2}tA^{-1}C^{\pm1}\big(q^\frac{1}{2}z\big)^{\pm1}\big)\Gamma_e\big((q p)^{\frac12 }t^{-1} A^{\pm1} C^{\pm1}(qz)^{\pm1}\big)\Gamma_e\big((q p)^{\frac12 }t^{-1} B^{\pm1} D^{\pm1}(qz)^{\pm1}\big)}{\Gamma\big((qz)^{\pm2}\big)}\\
\times\Gamma_e\big((qp)^{\frac12}t B^{-1}D^{\pm1} (qz)^{\pm1}\big)
\Gamma_e\big((q p)^{\frac12}t A C^{\pm1} (qz)^{\pm1}\big)\Gamma_e\big(q^{-1}A^2B^{-2}\big)\\
\times\Gamma_e\big(pq^2t^4\big)\Gamma_e\big(t^{-4}A^2B^{-2}z^2\big)\Gamma_e\big(z(qz)^{\pm1}\big)
\Gamma_e\big(pqt^4z^{-2}\big)\Gamma_e\big(q^{-1}t^{-4}A^2B^{-2}\big)\Gamma_e\big((qz)^{-1}(qz)^{\pm1}\big)\\
\times\Gamma_e\big(pq^2\big)\Gamma_e\big(pq^2t^4A^{-2}B^2z^{-1}(qz)^{\pm1}\big)\Gamma_e\big(t^{-4}z(qz)^{\pm1}\big)
\Gamma_e\big(A^{-2}B^2\big)T_{{\mathfrak J}_D}(qz) + \big\{z\leftrightarrow z^{-1}\big\},
\end{gather*}
which simplifies to
\begin{gather*}
\frac{\theta_p\big(pq^2t^2A^{-2}B^2\big)\theta_p\big((pq)^{-1}q^{-1}t^{-4}A^2B^{-2}\big)\Gamma_e\big(q^{-1}A^2B^{-2}\big)\Gamma_e\big(A^{-2}B^2\big)}
{\theta_p\big(t^{-2}\big)\theta_p\big(q^{-1}t^{-2}B^{-2}\big)\theta_p\big(q^{-1}t^{-2}A^2\big)\theta_p\big(pq^2t^4A^{-2}B^2\big)
\theta_p\big(q^{-1}t^{-2}AB^{-1}C^{\pm1}D^{\pm1}\big)}\\
\quad{}\times\frac{\theta_p\big((pq)^\frac{1}{2}t^{\pm1}BD^{\pm1}z\big)
\theta_p\big((pq)^\frac{1}{2}t^{\pm1}A^{-1}C^{\pm1}z\big)
\theta_p\big(q^{-1}t^{-4}\big)\theta_p\big(q^{-1}t^{-4}A^2B^{-2}z^2\big)}
{\theta_p\big(t^{-4}z^2\big)\theta_p\big(z^2\big)\theta_p\big(q^{-1}z^2\big)}T_{{\mathfrak J}_D}(z)\\
\quad{}+ \frac{\theta_p\big(pq^2t^2A^{-2}B^2\big)\theta_p\big((pq)^{-1}q^{-1}t^{-4}A^2B^{-2}\big)\Gamma_e\big(q^{-1}A^2B^{-2}\big)\Gamma_e\big(A^{-2}B^2\big)}
{\theta_p\big(t^{-2}\big)\theta_p\big(q^{-1}t^{-2}B^{-2}\big)\theta_p\big(q^{-1}t^{-2}A^2\big)
\theta_p\big(q^{-1}t^{-2}AB^{-1}C^{\pm1}D^{\pm1}\big)}\\
\quad{}\times\frac{\theta_p\big((pq)^\frac{1}{2}t^{-1}B^{\pm1}D^{\pm1}z\big)
\theta_p\big((pq)^\frac{1}{2}t^{-1}A^{\pm1}C^{\pm1}z\big)}
{\theta_p\big(qz^2\big)\theta\big(z^2\big)}T_{{\mathfrak J}_D}(qz) + \big\{z\leftrightarrow z^{-1}\big\} .
\end{gather*}
From the second term of \eqref{c7} we get
\begin{gather*}
\frac{\theta_p\big(pq^2t^2A^{-2}B^2\big)\theta_p\big((pq)^{-1}q^{-1}t^{-4}A^2B^{-2}\big)}
{\theta_p\big(t^{-2}\big)\theta_p\big(q^{-1}t^{-2}B^{-2}\big)\theta_p\big(q^{-1}t^{-2}A^2\big)\theta_p\big(pq^2t^4A^{-2}B^2\big)
\theta_p\big(q^{-1}t^{-2}AB^{-1}C^{\pm1}D^{\pm1}\big)}\\
{}\times\oint\frac{{\rm d}u}{4\pi i u} \frac{\Gamma_e\big((q p)^{\frac12 }t^{-1} A^{\pm1} D^{\pm1}u^{\pm1}\big)
\Gamma_e\big((q p)^{\frac12 }t^{-1} B^{\pm1} C^{\pm1}u^{\pm1}\big)}{\Gamma\big(u^{\pm2}\big)}\\
{}\times\Gamma_e\big((pq)^\frac12 t B^{-1} C^{\pm1}u^{\pm1}\big) \Gamma_e\big((pq)^\frac12 t A D^{\pm1}u^{\pm1}\big)
\Gamma_e\big((pq)^\frac{1}{2}t A^{-1}u^{\pm1}D^{\pm1}\big)
\Gamma_e\big((pq)^\frac{1}{2}t B u^{\pm1}C^{\pm1}\big)\\
{}\times\Gamma_e\big(AB^{-1}u^{\pm1}z^{\pm1}\big)
\Gamma_e\big((pq)^\frac{1}{2}q^\frac{1}{2}t B D^{\pm1}\big(q^\frac{1}{2}z\big)^{\pm1}\big)
\Gamma_e\big((pq)^{\frac{1}{2}}q^\frac{1}{2}tA^{-1}C^{\pm1}\big(q^\frac{1}{2}z\big)^{\pm1}\big)\\
{}\times\oint\frac{{\rm d}v}{4\pi i v} \frac{\Gamma_e\big((q p)^{\frac12 }t^{-1} A^{\pm1} C^{\pm1}v^{\pm1}\big)
\Gamma_e\big((q p)^{\frac12 }t^{-1} B^{\pm1} D^{\pm1}v^{\pm1}\big)}{\Gamma\big(v^{\pm2}\big)}\\
{}\times\frac{\theta_p\big((pq)^\frac{1}{2}q^{-1}t^{-3}B^{-1}z D^{\pm1}\big)
\theta_p\big((pq)^\frac{1}{2}q^{-1}t^{-3}A z C^{\pm1}\big)}
{\theta_p\big(z^2\big)\theta_p\big(t^4z^{-2}\big)}\\
{}\times\Gamma_e\big( A^{-1}B u^{\pm1} v^{\pm1}\big)\Gamma_e\big((qp)^{\frac12}t B^{-1}D^{\pm1} v^{\pm1}\big)
\Gamma_e\big((q p)^{\frac12}t A C^{\pm1} v^{\pm1}\big)T_{{\mathfrak J}_D}(v) + \big\{z\leftrightarrow z^{-1}\big\} .
\end{gather*}
Terms cancel in the integral over $u$ and what is left can be evaluated using the inversion formula as before which sets $v=z$, after some cancelations we get
\begin{gather*}
\frac{\Gamma_e\big(\big(AB^{-1}\big)^2\big)	\theta_p\big(pq^2t^2A^{-2}B^2\big)
\theta_p\big((pq)^{-1}q^{-1}t^{-4}A^2B^{-2}\big)}
{\theta_p\big(t^{-2}\big)\theta_p\big(q^{-1}t^{-2}B^{-2}\big)\theta_p\big(q^{-1}t^{-2}A^2\big)\theta_p\big(pq^2t^4A^{-2}B^2\big)
\theta_p\big(q^{-1}t^{-2}AB^{-1}C^{\pm1}D^{\pm1}\big)}\\
{}\!\times\!\frac{\theta_p\big((q p)^{\frac12 }t A^{-1} C^{\pm1}z\big)
\theta_p\big((q p)^{\frac12 }t B D^{\pm1}z\big)
\theta_p\big((pq)^\frac{1}{2}q^{-1}t^{-3}B^{-1}z D^{\pm1}\big)
\theta_p\big((pq)^\frac{1}{2}q^{-1}t^{-3}A z C^{\pm1}\big)}{\theta_p\big(z^2\big)\theta_p\big(t^4z^{-2}\big)}\\
{}\!\times\! T_{{\mathfrak J}_D}(z)+ \big\{z\leftrightarrow z^{-1}\big\} .
\end{gather*}
We compute the contribution from the last term in \eqref{c7}
\begin{gather*}
\oint\frac{{\rm d}u}{4\pi i u} \frac{\Gamma_e\big((q p)^{\frac12 }t^{-1} A^{\pm1} D^{\pm1}u^{\pm1}\big)
\Gamma_e\big((q p)^{\frac12 }t^{-1} B^{\pm1} C^{\pm1}u^{\pm1}\big)}{\Gamma\big(u^{\pm2}\big)}\\
{}\times\Gamma_e\big((pq)^\frac12 t B^{-1} C^{\pm1}u^{\pm1}\big)
\Gamma_e\big((pq)^\frac12 t A D^{\pm1}u^{\pm1}\big)
\Gamma_e\big((pq)^\frac{1}{2}tA^{-1}u^{\pm1}D^{\pm1}\big)
\Gamma_e\big((pq)^\frac{1}{2}t B u^{\pm1}C^{\pm1}\big)\\
{}\times\Gamma_e\big(A B^{-1}u^{\pm1}z^{\pm1}\big)
\Gamma_e\big((pq)^{\frac{1}{2}}t B D^{\pm1}z^{\pm1}\big)
\Gamma_e\big((pq)^{\frac{1}{2}}t A^{-1}C^{\pm1}z^{\pm1}\big)\\
{}\times\oint\frac{{\rm d}v}{4\pi i v} \frac{\Gamma_e\big((q p)^{\frac12 }t^{-1} A^{\pm1} C^{\pm1}v^{\pm1}\big)
\Gamma_e\big((q p)^{\frac12 }t^{-1} B^{\pm1} D^{\pm1}v^{\pm1}\big)}{\Gamma\big(v^{\pm2}\big)}\\
{}\times\Gamma_e\big( A^{-1}B u^{\pm1} v^{\pm1}\big)\Gamma_e\big((qp)^{\frac12}t B^{-1}D^{\pm1} v^{\pm1}\big)
\Gamma_e\big((q p)^{\frac12}t A C^{\pm1} v^{\pm1}\big)T_{{\mathfrak J}_D}(v) .
\end{gather*}
Integrals can be evaluated using the inversion formula which sets $v=z$ almost everything cancel and we get
\begin{gather*}
\Gamma_e\big(\big(AB^{-1}\big)^2\big)T_{{\mathfrak J}_D}(z) .
\end{gather*}
Dividing by this factor $\Gamma_e\big(\big(A B^{-1}\big)^2\big)$ and adding all contributions we get
\begin{gather*}
\frac{\theta_p\big(pq^2t^2A^{-2}B^2\big)
\theta_p\big((pq)^{-1}q^{-1}t^{-4}A^2B^{-2}\big)}
{\theta_p\big(t^{-2}\big)\theta_p\big(q^{-1}t^{-2}B^{-2}\big)\theta_p\big(q^{-1}t^{-2}A^2\big)
\theta_p\big(q^{-1}t^{-2}AB^{-1}C^{\pm1}D^{\pm1}\big)
\theta_p\big(q^{-1}A^2B^{-2}\big)}\\
{}\times \frac{\theta_p\big((pq)^\frac{1}{2}t^{-1}B^{\pm1}D^{\pm1}z\big)
\theta_p\big((pq)^\frac{1}{2}t^{-1}A^{\pm1}C^{\pm1}z\big)}{\theta_p\big(qz^2\big)\theta\big(z^2\big)}T_{{\mathfrak J}_D}(qz)\\
{}+ \frac{\theta_p\big(pq^2t^2A^{-2}B^2\big) \theta_p\big((pq)^{-1}q^{-1}t^{-4}A^2B^{-2}\big)}
{\theta_p\big(t^{-2}\big)\theta_p\big(q^{-1}t^{-2}B^{-2}\big)\theta_p\big(q^{-1}t^{-2}A^2\big)\theta_p\big(pq^2t^4A^{-2}B^2\big)
\theta_p\big(q^{-1}t^{-2}AB^{-1}C^{\pm1}D^{\pm1}\big)}\\
{}\times\frac{\theta_p\big((pq)^\frac{1}{2}t^{\pm1}BD^{\pm1}z\big)
\theta_p\big((pq)^\frac{1}{2}t^{\pm1}A^{-1}C^{\pm1}z\big)
\theta_p\big(q^{-1}t^{-4}\big)\theta_p\big(q^{-1}t^{-4}A^2B^{-2}z^2\big)}
{\theta_p\big(q^{-1}A^2B^{-2}\big)\theta_p\big(t^{-4}z^2\big)\theta_p\big(z^2\big)\theta_p\big(q^{-1}z^2\big)}T_{{\mathfrak J}_D}(z)\\
{}+ \frac{\theta_p\big(pq^2t^2A^{-2}B^2\big)
\theta_p\big((pq)^{-1}q^{-1}t^{-4}A^2B^{-2}\big)}
{\theta_p\big(t^{-2}\big)\theta_p\big(q^{-1}t^{-2}B^{-2}\big)\theta_p\big(q^{-1}t^{-2}A^2\big)\theta_p\big(pq^2t^4A^{-2}B^2\big)
\theta_p\big(q^{-1}t^{-2}AB^{-1}C^{\pm1}D^{\pm1}\big)}\\
{}\times\frac{\theta_p\big((q p)^{\frac12 }t A^{-1} C^{\pm1}z\big)
\theta_p\big((q p)^{\frac12 }t B D^{\pm1}z\big)
\theta_p\big((pq)^\frac{1}{2}t^3B D^{\pm1} z^{-1}\big)
\theta_p\big((pq)^\frac{1}{2}t^3 A^{-1} C^{\pm1} z^{-1}\big)}
{\theta_p\big(z^2\big)\theta_p\big(t^4z^{-2}\big)}\\
{}\times T_{{\mathfrak J}_D}(z)
+ \big\{z\leftrightarrow z^{-1}\big\} + T_{{\mathfrak J}_D}(z) .
\end{gather*}
Taking away overall factor of
\begin{gather*}
 \frac{\theta_p\big(pq^2t^2A^{-2}B^2\big)\theta_p\big((pq)^{-1}q^{-1}t^{-4}A^2B^{-2}\big)}
{\theta_p\big(t^{-2}\big)\theta_p\big(q^{-1}t^{-2}B^{-2}\big)\theta_p\big(q^{-1}t^{-2}A^2\big)
\theta_p\big(q^{-1}t^{-2}AB^{-1}C^{\pm1}D^{\pm1}\big) \theta_p\big(q^{-1}A^2B^{-2}\big)} ,
\end{gather*}
 we get
\begin{gather*}
\frac{\theta_p\big((pq)^\frac{1}{2}t^{-1}B^{\pm1}D^{\pm1}z\big)
\theta_p\big((pq)^\frac{1}{2}t^{-1}A^{\pm1}C^{\pm1}z\big)}
{\theta_p\big(qz^2\big)\theta\big(z^2\big)}T_{{\mathfrak J}_D}(qz)\\
{}+ \frac{\theta_p\big(q^{-1}t^{-4}\big)\theta_p\big(q^{-1}t^{-4}A^2B^{-2}z^2\big)
\theta_p\big((pq)^\frac{1}{2}t^{\pm1}BD^{\pm1}z\big)
\theta_p\big((pq)^\frac{1}{2}t^{\pm1}A^{-1}C^{\pm1}z\big)}{\theta_p\big(q^{-2}t^{-4}A^{2}B^{-2}\big)
\theta_p\big(t^{-4}z^2\big)\theta_p\big(z^2\big)\theta_p\big(q^{-1}z^2\big)}T_{{\mathfrak J}_D}(z)\\
{}+ \frac{\theta_p\big(q^{-1}A^2B^{-2}\big)\theta_p\big((pq)^\frac{1}{2}t^2B D^{\pm1} (t^{-1}z)^{\pm1}\big)
\theta_p\big((pq)^\frac{1}{2}t^2 A^{-1} C^{\pm1} (t^{-1}z)^{\pm1}\big)}
{\theta_p\big(q^{-2}t^{-4}A^{2}B^{-2}\big)\theta_p\big(z^2\big)\theta_p\big(t^4z^{-2}\big)}T_{{\mathfrak J}_D}(z)\\
{}+ \big\{z\leftrightarrow z^{-1}\big\}\nonumber\\
{}+ \frac{\theta_p\big(t^{-2}\big)\theta_p\big(q^{-1}t^{-2}B^{-2}\big)\theta_p\big(q^{-1}t^{-2}A^2\big)
\theta_p\big(q^{-1}t^{-2}AB^{-1}C^{\pm1}D^{\pm1}\big)
\theta_p\big(q^{-1}A^2B^{-2}\big)} {\theta_p\big(pq^2t^2A^{-2}B^2\big)
\theta_p\big((pq)^{-1}q^{-1}t^{-4}A^2B^{-2}\big)} T_{{\mathfrak J}_D}(z) .
\end{gather*}

To summarize, we have shown that
\begin{gather*}
T_{{\mathfrak J}_D}(v)\times_v \big(\big(T_{{\mathfrak J}_B, {\mathfrak J}_C,{\mathfrak J}_D}(w,u,v)\times_w C^{(0,0;A^{-1}B)}_{{\mathfrak J}_B}(w)\big)\\
\qquad{} \times_u\big(T_{{\mathfrak J}_B, {\mathfrak J}_C,{\mathfrak J}_D}(h,u,z)\times_h C^{(1,0;AB^{-1})}_{{\mathfrak J}_B}(h)\big)\big) ,
\end{gather*}
is proportional to
\begin{gather*}
 {\mathfrak D}_{{\mathfrak J}_D}^{{\mathfrak J}_B,(1,0;AB^{-1})} T_{{\mathfrak J}_D}(z) ,
\end{gather*}
where ${\mathfrak D}_{{\mathfrak J}_D}^{{\mathfrak J}_B,(1,0;AB^{-1})}$ is given by~(\ref{diuer}).

\section{Koornwinder limit of the trinion}\label{appendixD}

We compute here the index of the three punctured sphere in the Koornwinder limit. The only subtle issue is the evaluation of the integrals in the function~$H$~\eqref{gff}. The integral can be interpreted as index of two coupled ${\rm SU}(2)$ gauge theories with five flavors. The integrand does not have a good limit however by manipulating it using Seiberg dualities we can show that the limit is well defined and evaluate all the integrals. We will discuss the evaluation of this function here.

To evaluate $H$ we first perform Seiberg duality on $w_2$ splitting to five flavors in particular way, leading to ${\rm SU}(3)$ theory with five flavors
\begin{gather*}
\big((qp)^{\frac14}A^{-\frac12} t b z_1^{\pm1} , (qp)^{\frac14} A^{\frac12} a^{-1} t v_1^{\pm1} , (q p)^{\frac14} A^{-\frac12} b^{-1} z_2^{-1}\big) , \\
\big((q p)^{\frac12}\frac1{t^2} w_1^{\pm1} , (q p)^{\frac14} A^{\frac12}a v_2^{\pm1} , (q p)^{\frac14}A^{-\frac12}b^{-1} z_2\big) ,
\end{gather*} which leads to no $w_2$ flavors having negative powers of $p$. We have the mesons
\begin{gather*}
(q p)^{\frac34}A^{-\frac12}t^{-1} b z_1^{\pm1}w_1^{\pm1} , \ (q p)^{\frac34} A^{\frac12} a^{-1} t^{-1} w_1^{\pm1}v_1^{\pm1} , \ (q p)^{\frac34} t^{-2} A^{-\frac12} b^{-1} z_2^{-1} w_1^{\pm1} , \\
 (q p)^{\frac12} b a t z_1^{\pm1} v_2^{\pm1} , \ (q p)^{\frac12} t A v_2^{\pm1} v_1^{\pm1} , \ (q p)^{\frac12} A^{-1} b^{-2} , \\
 (q p) ^{\frac12} A^{-1} z_2 t z_1^{\pm1} , \ (q p)^{\frac12} a b^{-1} v_2^{\pm1} z_2 ^{-1} , \ (q p)^{\frac12} t b^{-1} a^{-1} z_2 v_1^{\pm1} , \end{gather*}
and the new charged fields are
\begin{gather*}
\square_{{\rm SU}(3)_{w_2}} \colon \ (p q)^{\frac16}A^{\frac13}b^{-\frac23} t^{\frac13}a^{-\frac23} z_2^{-\frac13} z_1^{\pm1} , \ (p q)^{\frac16}A^{-\frac23}b^{\frac13} t^{\frac13}a^{\frac13} z_2^{-\frac13} v_1^{\pm1} , \
( q p)^{\frac16} A^{\frac13} b^{\frac43} t^{\frac43} a^{-\frac23} z_2^{\frac23} , \\
\overline \square _{{\rm SU}(3)_{w_2 } } \colon \ (q p)^{\frac1{ 1 2} } t^{\frac23}A^{\frac16}a^{\frac23} b^{-\frac13}z_2^{\frac13}w_1^{\pm1} , \
(q p) ^{\frac13} A^{-\frac13} b^{-\frac13} t^{-\frac43} a^{-\frac13} z_2^{\frac13} v_2^{\pm1} , \ ( q p )^{\frac13} A^{\frac23} b^{\frac23} t^{-\frac43} a^{\frac23} z_2^{-\frac23} ,
\end{gather*}
 and all of these fields have a good limit when $p$ goes to zero and the fugacities are scaled. We note that some of the mesons charged under $w_1$ form mass terms with some of the quarks and after the first Seiberg duality we have four flavors of $w_1$
\begin{gather*}
 \big( (q p )^{\frac14} A^{\frac12}b z_2 , (q p) ^{\frac1{ 1 2 }} A^{\frac16} t^{\frac23} a^{\frac23} b^{-\frac13} z_2^{\frac13} \big(w_2^j\big)^{-1} \big) , \\
 \big((q p)^{\frac14} A^{\frac12} b z_2^{-1} , (q p)^{\frac14} A^{-\frac12} a^{-1} v_2^{\pm1} , (q p)^{\frac34} A^{-\frac12} t^{-2} b^{-1} z_2^{-1} \big) ,
\end{gather*}
 for which we perform another duality operation. After the duality the $p$ vanishing limit is well defined for all fields. Let us write the fields surviving after scaling
\begin{gather*}
(p q)^{\frac16}A^{\frac13}b^{-\frac23} t^{\frac13}a^{-\frac23} z_2^{-\frac13} z_1^{\pm1}w_2^j , \ (p q)^{\frac16}A^{-\frac23}b^{\frac13} t^{\frac13}a^{\frac13} z_2^{-\frac13} v_1^{\pm1} w_2^j , \\
(q p)t^{-2} , \ (q p)^{\frac13} A^{-\frac13} t^{-\frac43} b ^{-\frac13} a^{-\frac13} z_2^{\frac13} v_2^{-1} \big(w^j_2\big)^{-1} , \
 (q p)^{\frac13} A^{-\frac13} t^{\frac23} b ^{-\frac13} a^{-\frac13} z_2^{\frac13} v_2^{-1} \big(w^j_2\big)^{-1} , \
a b^{-1} t w_1^{\pm1} ,\\
( q p)^{ \frac1 2 }a b^{-1} v_2^ {-1} z_2^{-1} , \ ( q p)^{\frac12} b^{-1} t^{-1} A^{-1} w_1^{\pm1} a^{-1} ,\\
(q p)^{\frac12} a^{-1} b z_2 v_2 , \ (q p)^{\frac12} z_2^{-1} t^{-1} w_1^{\pm1} v_2^{-1} , \
b t a^{-1} w_1^{\pm1} , \ (q p)^{\frac12} t v_2 A v_1^{\pm1} , \ (q p)^{\frac12} a b t v_2 z_1^{\pm1} ,
\end{gather*}
and we can evaluate the index of the three punctured sphere to be \eqref{krlty}.

\subsection*{Acknowledgments}
We would like to thank Hee-Cheol Kim, S.~Ruijsenaars, Cumrun Vafa, and Gabi Zafrir for relevant discussions. The research was supported by Israel Science Foundation under grant no. 1696/15 and by I-CORE Program of the Planning and Budgeting Committee.

\pdfbookmark[1]{References}{ref}
\LastPageEnding

\end{document}